\documentclass[11pt]{article}

\usepackage{color}

\usepackage{authblk}
\usepackage{url}
\usepackage{graphicx}
\usepackage{graphics,epsfig}
 \usepackage{amssymb}
 \usepackage{ulem}
 \usepackage{multirow}
  \usepackage{color}
  \usepackage{subfigure}
  \usepackage{subfig}
 \usepackage{amsmath}
\usepackage[margin=1in]{geometry}
\usepackage[numbers,sort&compress]{natbib}
\usepackage{enumitem}
\usepackage{comment}

\newcommand{\bfl}{\begin{flushleft}}
\newcommand{\efl}{\end{flushleft}}
\newcommand{\bea}{\begin{eqnarray}}
\newcommand{\eea}{\end{eqnarray}}
\newcommand{\be}{\begin{equation}}
\newcommand{\ee}{\end{equation}}
\newcommand{\ben}{\begin{enumerate}[itemsep=0pt,parsep=0pt]}
\newcommand{\een}{\end{enumerate}}
\newcommand{\bi}{\begin{itemize}}
\newcommand{\ei}{\end{itemize}}
\newcommand{\tben}{\begin{enumerate}[itemsep=0.0in,parsep=0.0in]}
\newcommand{\teen}{\end{enumerate}}

\newcommand{\gev}{~\mbox{GeV}}

\newcommand{\vp}{\varphi}

\newcommand{\sss}{{\scriptscriptstyle \Sigma}}
\newcommand{\CM}{{\mathcal M}}
\newcommand{\CK}{{\mathcal K}}

\newcommand{\gsim}{\lower.7ex\hbox{$\;\stackrel{\textstyle>}{\sim}\;$}}
\newcommand{\lsim}{\lower.7ex\hbox{$\;\stackrel{\textstyle<}{\sim}\;$}}

\renewcommand{\em}{\it}

\def\bec{\begin{center}}
\def\eec{\end{center}}
\def\beq{\begin{equation}}
\def\eeq{\end{equation}}
\def\fr{\frac}

\def\vp{\varphi}

\title{How Well Can We Really Determine the Scale of Inflation?}
\author{Ogan \"Ozsoy}
\author{Kuver Sinha}
\author{Scott Watson}

\affil{Department of Physics, Syracuse University, Syracuse, NY 13244, USA}

\date{\today}
\begin{document}
\maketitle

\begin{abstract}
A detection of primordial B-modes has been heralded not only as a smoking gun for the existence of inflation, but also as a way to establish the scale at which inflation took place.
In this paper we critically reinvestigate the connection between a detection of primordial gravity waves and the scale of inflation.  We consider whether the presence of additional fields and 
non-adiabaticity during inflation may have provided an additional source of  primordial B-modes  competitive with those of the quasi-de Sitter vacuum.  In particular, we examine whether the additional sources could provide the dominant signal, which could lead to a misinterpretation of the scale of inflation.  In light of constraints on the level of non-Gaussianity coming from Planck we find that only hidden sectors with strictly gravitationally strength couplings provide a feasible mechanism.  The required model building is somewhat elaborate, and so we discuss possible UV completions in the context of Type IIB orientifold compactifications with RR axions.  We find that an embedding is possible and that dangerous sinusoidal corrections can be suppressed through the compactification geometry.  Our main result is that even when additional sources of primordial gravity waves are competitive with the inflaton, a positive B-mode detection would still be a relatively good indicator of the scale of inflation.  This conclusion will be strengthened by future constraints on both non-Gaussianity and CMB polarization. 
\end{abstract}
\thispagestyle{empty}

\newpage

\section{Introduction}
A positive detection of B-mode polarization in the Cosmic Microwave Background (CMB) -- if identified as being of primordial origin -- has been argued to provide {\em smoking gun} evidence for the existence of inflation \cite{Baumann:2008aq}. It has been further argued that the signal would provide us with the scale at which inflation took place.  Given the current and projected sensitivity of polarization experiments \cite{Baumann:2008aq}, a positive detection of primordial B-modes would then imply inflation occurred near the GUT scale, or slightly below.  Indeed, if the results from BICEP2 \cite{Ade:2014xna} are confirmed, this would be the first direct evidence for physics beyond the Standard Model at a scale nearly a billion times that probed at the Large Hadron Collider (LHC).  

In this paper we revisit the question;  {\em Does a  
detection of primordial B-modes necessarily provide us with the scale of inflation}? In \cite{Senatore:2011sp} it was 
argued that the answer is no. In that paper, the authors considered additional sources of gravity waves arising from non-adiabaticity and particle production during inflation and claimed that in some cases this source of B-modes could exceed those
coming from the quantum fluctuations of the quasi-de Sitter background.  Related ideas have appeared in 
\cite{Barnaby:2012xt,Barnaby:2012tk,Barnaby:2011qe,Barnaby:2011vw,Barnaby:2009dd,Cook:2013xea,Cook:2011hg,Caprini:2014mja,Sorbo:2011rz,Carney:2012pk,Pajer:2013fsa,Fedderke:2014ura}, although the primary focus of these papers was different. In this paper we will review both approaches and explicitly demonstrate their relation for the case of on-shell particle production.  In many of these works it was also pointed out that the same effects leading to a significant level of gravity waves would also lead to a substantial level of equilateral type non-Gaussianity (NG) -- a prediction that was important for Planck.  Utilizing the current Planck data \cite{Ade:2013ydc} we can now revisit these models utilizing the constraints on the level of equilateral type NG $f^{\mbox{\tiny equil}}_{\mbox{\tiny NL}}<-42 \pm 75$.  Using this constraint, and demanding successful inflation and self consistent model building, in this paper we examine these models to see if particle production can lead to a competitive source for primordial B-modes.

We first consider the case of an inflaton directly coupled to spectator fields.  This captures models with on-shell particle production such as Trapped Inflation and Moduli Trapping \cite{Kofman:2004yc,Watson:2004aq,Green:2009ds}, and we also consider production of pseudo scalar and gauge fields during inflation \cite{Sorbo:2011rz}.  In all of these models we find that the direct coupling typically leads to a high level of NG, rendering these alternatives for generating primordial B-modes irrelevant.  We next consider the production of spectator fields with gravitational coupling to the inflaton sector \cite{Barnaby:2012xt,Cook:2013xea}. 
Because of the suppressed couplings, in some cases these models can lead to a lower level of NG and an alternative B-mode source is possible. However, additional constraints from back-reaction and isocurvature perturbations severely restrict the parameter space.  We identify the most promising case as gauge field production resulting from a tachyonic and time dependent mass term resulting from the interaction of the gauge field with an additional spectator scalar field (not the inflaton).  Given the elaborate nature of this model, after constraining the parameter space we turn to the question of UV completing the model.  We construct an inflationary sector utilizing Axion Monodromy \cite{McAllister:2008hb, McAllister:2014mpa, Flauger:2009ab, Kaloper:2008fb, Marchesano:2014mla, Hebecker:2014eua, Kenton:2014gma}, and then we realize the additional spectator fields needed within the framework of $O3/O7$ orientifold compactifications of Type IIB string theory.  
We find that a UV embedding can be realized in the weakly coupled string theory if the compactification volume is taken parametrically larger than the Planck scale and if the axion decay constant is sub-Planckian.
Our embedding also demonstrates that dangerous sinusoidal corrections to the gauge field production models can be suppressed through the compactification geometry.

The remainder of the paper is as follows.  In the next section we review both the classical and quantum production of gravity waves during inflation. This section is primarily to establish notation and to address a few subtle points in the literature regarding the production of gravity waves from on-shell particle production.  In Section \ref{direct_coupling_section}, we consider the case of particle production of fields directly coupled to the inflationary sector.  In Section \ref{section4}, we consider the gravitationally coupled case.  Unlike the direct coupling case, we find that some of these models do result in B-modes although the parameter space is severely restricted\footnote{The authors of \cite{Ferreira:2014zia} have argued for slightly different constraints than we find for these models. In particular, their analysis of the behavior of the hidden sector at Hubble radius crossing leads to stronger constraints from the curvature power spectrum and bispectrum than those found in the existing literature.  If confirmed, this would strengthen our conclusions.  We refer the reader to their paper for details.}.  In the remainder of the paper, we consider the UV completion of these models.  First, in Section \ref{UVchallenges} we review the relevant details of Type IIB orientifold compactifications and their role in Inflationary model building.  Then in Section \ref{stringy} we present an explicit model including the particle production and establish the model building constraints.  In the final section we conclude.   In appendix A we list a number of  concerns for string model building with moduli stabilization and their possible resolutions.

\section{Gravity Waves from Inflation and Particle Production}
In this section we review the general formalism for establishing the amount of gravity waves produced during inflation from both quantum fluctuations of the metric and classical sources from particle production events during inflation. For readers familiar with these types of calculations this section may be skipped, however it does serve to set our notation and conventions.

Gravity waves produced during inflation can perturb the homogeneous and isotropic background metric. These tensor fluctuations are described by the metric 
\be
ds^2 = a(\tau)^2[-d\tau^2 + (\delta_{ij}+h_{ij})dx^{i}dx^{j}],
\ee
where Latin indices denote spatial co-ordinates\footnote{We will follow metric signature $(-,+,+,+)$ and work with the reduced Planck mass $m_p=2.44 \times 10^{18}$ GeV.}, $h_{ij}$ is the transverse ($\partial_{i}h_{ij}=0$) and traceless ($h_{ii}=0$) metric perturbation and we work in conformal time with $a=-1/(H\tau)$ for quasi-dS. 

The mode equation for gravity waves in the cosmological background (working in Fourier space) is  
\be \label{modeqn}
\bar{h}^{\prime \prime}_{ij} +\left( k^2-\frac{a^{\prime\prime}}{a}\right)\bar{h}_{ij} = \frac{2}{m_p^2}~a~T^{TT}_{ij},
\ee
where we introduced canonical modes $\bar{h}_{ij}=a(\tau) h_{ij}$ and $T^{TT}_{lm}$ is the transverse and traceless components of the stress energy tensor for any sources which are present. The transverse, traceless components of the stress tensor can be obtained by introducing the projector
$\Pi_{ij}^{\; \; lm}=P_i^{\;l} P_j^{\;m}-\frac{1}{2} P_{ij} P^{lm}$ where $P_{ij}=\delta_{ij}-k_i k_j /k^2$ so that $T^{TT}_{ij}= \Pi_{ij}^{\; \; lm} T_{lm}$ (cf. \cite{Misner:1974qy}).

We can formally solve \eqref{modeqn} to find
\be \label{grav_source}
\bar{h}_{ij}(\vec{k},\tau)=\frac{2}{m_p^2} \int d\tau^\prime \; G_k(\tau,\tau^\prime)~ a(\tau^{\prime})~ T^{TT}_{ij}(\vec{k},\tau^\prime),
\ee
where $G_k(\tau,\tau^\prime)$ is the Green function satisfying the source free version of \eqref{modeqn} with appropriate boundary conditions.
For the quasi-dS background we find
\be \label{greeny}
G_k(\tau,\tau^\prime)=\frac{k(\tau^\prime-\tau)  \cos \left( k(\tau^\prime - \tau) \right)-(1+k^2 \tau^\prime \tau  ) \sin \left( k(\tau^\prime - \tau) \right) }{k^3 \tau \tau^\prime} \; \Theta(\tau-\tau^\prime),
\ee
where $\Theta(\tau-\tau^\prime)=0$ for $\tau<\tau^\prime$ signaling that the source only produces gravity waves after its creation.  This expression along with a source in \eqref{grav_source} then allows us to find the resulting gravitational radiation. 

\subsection{Quantum Vacuum Fluctuations and Gravity Waves}
For inflationary vacuum fluctuations and in the absence of sources ($T_{l m}=0$) equation \eqref{modeqn} can be easily solved (see e.g. \cite{Mukhanov:1990me} or \cite{Baumann:2014nda}) and one finds that the 
inflationary background generates a nearly scale invariant spectrum of gravitation waves. We can relate the correlation function to the tensor power spectrum for each helicity as,
\be \label{tensors}
\fr{1}{a^2}\langle \bar{h}^s_{ij}(\vec{k})~ \bar{h}^{s^\prime}_{ij}(\vec{k}^{\prime})  \rangle =  (2\pi)^3 \delta^{(3)}(\vec{k}+\vec{k}^\prime) \delta^{ss^\prime} {\cal P}_h,
\ee
where $s$ refers to the polarization.  The dimensionless tensor power-spectrum resulting from quantum vacuum fluctuations of the graviton is then
\be \label{power}
\Delta^2_t(k) = 2 \cdot \frac{k^3}{2 \pi^2} \cdot {\cal P}_h=\frac{8}{m_p^2}\left( \frac{H}{2 \pi} \right)^2
\ee
evaluated at $k=aH$ and the tilt of the spectrum is $n_t = d \ln {\cal P}_h / d \ln k$. 
In the absence of any other primordial sources of gravity waves, a measurement of the tensor-to-scalar ratio (along with existing measurements of the scalar power spectrum) then 
allow us to determine the scale of inflation $H_I$ through \eqref{power}.  In terms of the scalar power spectrum ${\Delta}_s^2$ we can then define the tensor to scalar ratio
\bea
r &\equiv& \frac{\Delta^2_t}{\Delta^2_s}.
\eea
where near-term and future experiments can be optimistically expected to probe as low as $r\simeq 10^{-3}$ \cite{Baumann:2008aq}. 
Using the COBE normalization $\Delta^2_s =2.5 \times 10^{-9}$ this can be re-expressed as a determination of the scale of inflation (cf. \cite{Baumann:2014nda})
\be
H_I=3 \times 10^{-5} \left( \frac{r}{0.1}\right)^{1/2} m_p.
\ee
Thus, given an observation of $r$ and knowledge that the only source of gravity waves resulted from primordial vacuum fluctuations, we can determine the scale of inflation $H_I \sim V^{1/2} / m_p$ \footnote{In \cite{Ashoorioon:2013eia}, the issue of whether the choice of the Bunch-Davies vacuum as the initial condition for perturbations is important or not was discussed.}.

\subsection{Gravity Waves from Particle Production during Inflation}
We will first consider gravity wave sources from scalar field production during inflation and later generalize this to vector fields. To calculate the effect of the produced particles on gravity wave production we note that the contribution of the particles to the spatial part of the stress tensor will be of the form
$T_{ij}= \partial_i \chi \partial_j \chi + \delta_{ij} (\ldots)$.  This implies that in Fourier space the transverse, traceless source is the convolution \cite{Dufaux:2007pt}
\be
T^{TT}_{ij}(k,\tau) =\Pi_{ij}^{\; \; lm}(k) \int \frac{d^3 p}{(2\pi)^{3/2}} \; p_l  \, p_m  \, \chi(p,\tau)\, \chi(k-p,\tau).
\ee
Using this result, along with \eqref{grav_source} we can construct the two point, equal time correlator 
\bea \label{2pt}
\langle \bar{h}_{ij}(k,\tau) \, \bar{h}_{ij}^*(k^\prime,\tau) \rangle &=& \frac{4}{m_p^4} \int d\tau' a(\tau') \; G_k(\tau,\tau')  \int d\tau''a(\tau'') \; G_{k^\prime}(\tau,\tau^{\prime \prime}) \Big\langle T^{TT}_{ij}(k,\tau^\prime) 
T^{TT*}_{ij}(k^\prime,\tau^{\prime\prime})  \Big\rangle \nonumber \\
&=& \frac{4}{m_p^4} \int d\tau' \; \frac{G_k(\tau,\tau')}{a(\tau')}  \int d\tau'' \; \frac{G_{k^\prime}(\tau,\tau'')}{{a(\tau'')}}\Pi_{ij}^{\; \; lm}(k) \Pi_{ij}^{\; \; no}(k^\prime)\int \frac{d^3 p}{(2\pi)^{3/2}}   \int \frac{d^3 p^\prime}{(2\pi)^{3/2}} \, \nonumber \\
&\times&
 p_l  \, p_m~ p_n^\prime  \, p'_o \,  \, \Big\langle \hat{\chi}(p,\tau^\prime)\, \hat{\chi}(k-p,\tau^\prime) \hat{\chi}^*(p^\prime,\tau^{\prime \prime})\, \hat{\chi}^*(k^\prime-p^\prime,\tau^{\prime \prime}) \Big\rangle.\;\;\;\;\;\;\;\;
\eea
where we have introduced the canonical field $\hat{\chi}= a(\tau) \chi$.  

As discussed in \cite{Dufaux:2007pt} if we now assume that the fields are well approximated by statistically homogeneous, random Gaussian fields than the four-point function can be written in terms 
of two-point functions by Wick's theorem\footnote{The authors of \cite{Dufaux:2007pt} pointed out that this is a good approximation at both the beginning and end of inflationary preheating and is in good agreement with lattice simulations. Here we consider these events during inflation and will work within the same approximation.  We will see that any strong coupling of the spectator field $\chi$ will tend to generate a large level of non-Gaussianity making this approximation justified given existing CMB constraints.}.  Using that
\be \label{chi2pt}
\Big\langle \hat{\chi}(p,\tau^\prime)\, \hat{\chi}^*(p^\prime,\tau^{\prime \prime})  \Big\rangle = f(p,\tau^\prime, \tau^{\prime \prime}) \delta(p-p^\prime),
\ee
for statistically homogeneous and isotropic fields and keeping only the connected pieces of the correlator we have
\begin{multline}
\Big\langle \hat{\chi}(p,\tau^\prime)\, \hat{\chi}(k-p,\tau^\prime) \hat{\chi}^*(p^\prime,\tau^{\prime \prime})\, \hat{\chi}^*(k^\prime-p^\prime,\tau^{\prime \prime}) \Big\rangle_{connected} =\nonumber \\
\delta(k-k^\prime) f(p,\tau^\prime, \tau^{\prime \prime}) f(k-p,\tau^\prime, \tau^{\prime \prime}) \left[ \delta(p^\prime-p) + \delta(p^\prime-k+p) \right]
\end{multline}

Using this result and the property of the projectors that $\Pi_{ij}^{\; \; lm} \Pi_{ij}^{\; \; no} p_l p_m p_n p_o=\Pi^{lmno} p_l p_m p_n p_o=p^4 \sin^4(\theta) /2$ where $\theta$ is the angle between $k$ and $p$,  we can perform one of the momentum integrals in \eqref{2pt} and we find

\begin{multline} \label{gravity2pt}
\langle h_{ij}(k,\tau) \, h^*_{ij}(k^\prime,\tau) \rangle
= \\
\delta(k-k^\prime) \frac{2}{m_p^4} \int \frac{{d\tau^\prime}}{a^2(\tau^{\prime})} \; G_k(\tau,\tau^\prime)  \int \frac{d\tau^{\prime \prime}}{a^2(\tau^{\prime \prime})} \; G_{k^\prime}(\tau,\tau^{\prime \prime})
\int \frac{d^3 p}{(2\pi)^{3}}
p^4 \sin^4(\theta)  f(p,\tau^\prime, \tau^{\prime \prime}) f(k-p,\tau^\prime, \tau^{\prime \prime})
\end{multline}

It remains to determine the functions in the two-point correlator \eqref{chi2pt}.  
We are interested in cases where quanta of the $\hat{\chi}$ field become excited due to the interaction with the inflaton and particle production.
To calculate this contribution to \eqref{chi2pt} we will follow the treatment in \cite{Kofman:1997yn}.  

The equation of motion for the canonical field is
\be \label{wsq1}
{\chi}^{\prime \prime}+\omega^2(\tau) {\chi}=0,
\ee
where we make the change of notation $\hat{\chi} \rightarrow \chi$ for simplicity and where
\be \label{wsq2}
\omega^2(\tau) = k^2 +a^2 m^2_{eff}(\tau) - a^2 \Delta.
\ee
where $\Delta \sim a^{\prime \prime} /a$ is typically negligible compared to the effective time-dependent mass $m_{eff}$.
The WKB solution to this equation 
\be
\chi(k,\tau)= \frac{1}{\sqrt{2 \omega_k}} \left(  \alpha_k(\tau) e^{-i \int^\tau \omega_k(\tilde{\tau}) d\tilde{\tau}} + \beta_k(\tau) e^{i \int^\tau \omega_k(\tilde{\tau}) d\tilde{\tau}} \right)
\ee
is valid as long as $\omega^\prime < \omega^2$ and all higher order adiabatic invariants remain small.  The condition that initially there are 
no quanta of the field present requires $\alpha=1$ and $\beta=0$, i.e. only positive frequency modes are present. 
When the adiabatic conditions fails, particle production results, and the Bogolyubov coefficients above give the mode mixing that occurs due to the time dependence of the system.
The creation and annihilation operators after production can be expanded in terms of the initial creation $\hat{a}_k^\dagger$ and annihilation operators $\hat{a}_k$ of the field as
\bea
\hat{b}_k(\tau)=\alpha_k(\tau) \hat{a}_k + \beta^*_k(\tau) \hat{a}_{-k}^\dagger, \;\;\;\;\;\;\;\;\;\;
\hat{b}_k^\dagger(\tau)=\alpha_k(\tau) \hat{a}_k + \beta^*_k(\tau) \hat{a}_{-k}^\dagger, 
\eea
where although $\hat{a}_k$ annihilates the vacuum initially, $\hat{b}_k$ does not, and if the system returns to adiabatic evolution the number density of particles produced is $n_k \sim |\beta_k|^2$.
Proper renormalization requires that one normal orders the correlators with respect to the $\hat{b}_k$ basis and then uses that $\hat{a}_k$ annihilates the vacuum to find the surviving terms (see e.g. \cite{Cook:2011hg,Barnaby:2012xt}).  Expanding the field and performing the normal ordering we find that the unknown function in \eqref{chi2pt} is
\begin{multline}
f(p,\tau^\prime, \tau^{\prime \prime})=\frac{1}{2 \sqrt{\omega_p(\tau^\prime) \omega_p(\tau^{\prime \prime})}}
\left\{
\alpha_p(\tau^\prime) \beta^*_p(\tau^{\prime \prime}) e^{-i \int^{\tau^{\prime }}  \omega_k(\tilde{\tau}) d\tilde{\tau} -i \int^{\tau^{\prime \prime}} \omega_k(\tilde{\tau}) d\tilde{\tau} }
\right. \\ \left. \beta_p(\tau^\prime) \alpha^*_p(\tau^{\prime \prime}) e^{i \int^{\tau^{\prime }}  \omega_k(\tilde{\tau}) d\tilde{\tau} +i \int^{\tau^{\prime \prime}} \omega_k(\tilde{\tau}) d\tilde{\tau} }
+\beta_p(\tau^\prime) \beta^*_p(\tau^{\prime \prime}) e^{-i \int^{\tau^{\prime \prime}}_{\tau^\prime} \omega_k(\tilde{\tau}) d\tilde{\tau}}
+ \beta^*_p(\tau^\prime)\beta_p(\tau^{\prime \prime})  e^{i \int^{\tau^{\prime \prime}}_{\tau^\prime} \omega_k(\tilde{\tau}) d\tilde{\tau}} \right\}
\end{multline}
We then use this expression in \eqref{gravity2pt} to find the amount of gravitational radiation.  
However, as shown in \cite{Barnaby:2012xt} the arguments of the exponentials above lead to rapidly oscillating phases and don't give a significant contribution to the final correlator \eqref{gravity2pt}.  Neglecting the phases, using the result above in \eqref{gravity2pt} and keeping only the leading terms we find

\begin{multline} \label{result}
\langle h_{ij}(k,\tau) \, h^*_{ij}(k^\prime,\tau) \rangle
= \\
\delta(k-k^\prime) \frac{2}{m_p^4} \int \frac{{d\tau^\prime}}{a^2(\tau^{\prime})} \; G_k(\tau,\tau^\prime)  \int \frac{d\tau^{\prime \prime}}{a^2(\tau^{\prime \prime})} \; G_{k^\prime}(\tau,\tau^{\prime \prime})
\int \frac{d^3 p}{(2\pi)^{3}}
p^4 \sin^4(\theta)  \frac{|\beta_p|^2 (|\alpha_p|^2+|\beta_p|^2)}{2 \omega_p(\tau^\prime) \omega_p(\tau^{\prime \prime}) } + \ldots,
\end{multline}
where the missing terms are sub-leading and we refer to \cite{Barnaby:2012xt} for a more detailed discussion.    The result \eqref{result} will allow us in the remainder of the paper to connect particle production and non-adabaticity during inflation with the generation of gravitational waves.  Given the Green function \eqref{greeny}, we simply calculate the Bogolyubov coefficients for a given model and this gives us the associated gravity waves produced via \eqref{result}. Given this contribution to the tensor power spectrum we can then compare to the vacuum source in \eqref{power}
to determine if particle production can lead to a larger signal.

\section{Particle Production Mechanisms with Direct Inflaton Coupling \label{direct_coupling_section}}
Any contribution to the production of gravitational waves during inflation, if competitive to vacuum fluctuations, could obstruct the use of observations to determine both the scale of inflation and whether the waves are of classical or quantum mechanical origin -- with quasi-dS fluctuations exemplifying the latter. In this section we consider the gravitational waves resulting from the production of fields directly coupled to the inflaton, and establish constraints for whether such effects can be competitive. 

Inflation models in best agreement with existing data are necessarily sensitive to high energy (UV) physics.
Thus, consistent model building requires these models to be embedded in a UV complete theory, with string theories currently providing the most developed approach.
String theories come with additional fields, strings and branes, and the importance of these degrees of freedom on the inflation process has been 
demonstrated in a number of contexts (see \cite{Baumann:2014nda} for a review).  Among the anticipated effects, if these states couple to the inflaton during inflation this can lead to particle production, which in some cases may be expected
to generate a large background of gravitational waves.

Following \cite{Gubser:2003vk} (see also \cite{Senatore:2011sp}) our starting point is the action 
\be
S= \int \sqrt{-g} \left[ \frac{1}{2} m_p^2 R - \frac{1}{2} (\partial \vp)^2 - V(\vp) \right]+S_p+S_s+S_{int},
\ee
where for particle sources we have
\be \label{source1}
S_p=-\sum_p \int d^4x \int d\tau \, \delta^{(4)}\left( x^\mu - x^{\mu}_p(\tau) \right) \theta(t-t_p) \; m(\vp) \sqrt{-g_{\mu \nu}(\tau) \frac{dx^\mu}{d\tau} \frac{dx^\nu}{d\tau}},
\ee
and the mass $m(\vp)$ of the particle depends on the inflaton (and so co-ordinate time $t$), $t_p$ is the time at which a particle is created, and the argument of the square root is given by the world line trajectory of the particle and so depends on the proper time $\tau$.  Similarly for string sources one has 
\be \label{source2}
S_s=-\sum_s \int d^4x \int d^2\sigma \, \delta^{(4)}\left( x^\mu - x^{\mu}_s(\sigma) \right) \theta(t-t_s) \; T(\vp) \sqrt{- \det \left( g_{\mu \nu}(\sigma) \frac{dx^\mu}{d\sigma^\alpha} \frac{dx^\nu}{d\sigma^\beta} \right)} 
\ee
where the string tension $T(\vp)$ can depend on the inflaton and $\sigma = (\tau, \sigma_1)$ are the 
induced coordinates on the string world volume and $t_s$ is the time of string production.  $S_{int}$ accounts for any interactions between the inflationary sector and the particles and strings.

In the rest of the section, we go through several examples of how couplings in \eqref{source1} can lead to particle and string production during inflation.  Our examples will cover the cases of a scalar and pseudo-scalar inflaton, producing scalar or vector particles. We will then go on to describe gravity wave production in these models, and finally constraints from non-Gaussianity and back-reaction.

\subsection{Scalar Production During Inflation}
The time dependent mass and tension appearing in \eqref{source1} and \eqref{source2} can lead to interesting cosmological implications.
In models of moduli trapping \cite{Kofman:2004yc,Watson:2004aq,Cremonini:2006sx,Greene:2007sa} the particle production resulting from the scalar's time dependence was shown to lead to dynamical stabilization of moduli (massless scalars) that would otherwise have little or no potential and remain unstablized.  When identified as the inflaton, it was shown in \cite{Green:2009ds,Dong:2010in} that when accounting for the effects of particle production one can obtain slow-roll inflation from potentials that would otherwise not satisfy the slow-roll conditions.
Taking the scalar in \eqref{source1} and \eqref{source2} to be the inflaton we can capture the dynamics of this production through an effective interaction
\be \label{scalar_prod}
\mathcal{L}_{int} \, = \, g^2 (\varphi - \varphi_0)^2 \chi^2.
\ee
Here $\varphi$ is the inflaton and $\chi$ the spectator field to be produced.  Although this interaction is much simpler than what one might expect from \eqref{source1} and \eqref{source2}, since we are treating $\chi$ as a simple scalar, it was shown in \cite{Gubser:2003vk} that this also provides an adequate description of string and brane production within the low-energy effective theory.

{\em How generic is such an interaction?} Interactions captured by \eqref{scalar_prod} generically occur within the context of string theory and M-theory model building.  Common examples include the presence of new light states (here represented as $\chi$) as the size of a compact dimension or internal cycle (parametrized by $\varphi$) shrink and the symmetries of the theory become enhanced \cite{Watson:2004aq,Cremonini:2006sx} -- i.e. there is an inverse string Higgs effect.  Other examples include when D-branes become coincident, as in models of brane inflation, and new light states appear\footnote{These new states correspond to open strings stretching between the branes becoming light.} \cite{Kofman:2004yc}, or near locations in field space associated with changes in topology \cite{Greene:2007sa}.  In these and other cases the interaction is effectively captured by \eqref{scalar_prod} where far from the location $\varphi=\varphi_0$ the quanta of $\chi$ can be quite heavy and so don't effect the dynamics.  However, as the inflaton $\varphi$ approaches $\varphi_0$ quanta of the $\chi$ field can become excited leading to on-shell\footnote{In this paper we will focus on on-shell production, whereas the interaction \eqref{scalar_prod} can also lead to important off-shell (virtual) effects as discussed first in \cite{Silverstein:2003hf}.  Which effect is dominant depends on whether the theory is in the strong or weak coupling regime as discussed in \cite{Kofman:2004yc}.}  particle production\footnote{This is analogous to Schwinger pair production in a strong electromagnetic field.}. 

Given this motivation we are now interested in determining the amount of gravitational waves that could be generated during the creation process, while still allowing for successful inflation and being consistent with bound on non-Gaussianity.  We first emphasize that scalar field waves do not produce an appreciable amount of gravitational radiation \cite{Dufaux:2007pt}.   Instead, here the expectation of gravity waves comes from both the creation process and the existence of the particles following creation as they provide a classical source in \eqref{grav_source}.
We emphasize that the created particles are inhomogenously distributed,  on-shell, and are not perturbations\footnote{This explains why it is consistent to use the linearized equation for the graviton in \eqref{grav_source}, whereas we will see that the sources will be quadratic in the created fields.} \cite{Dufaux:2007pt}.  Moreover, the creation process is non-perturbative and cannot be described within standard methods of linearlized perturbation theory\footnote{It was for many of these reasons that in \cite{Carney:2012pk} it was shown that within perturbation theory the production of gravitational waves would be negligible (see also \cite{Dufaux:2007pt}).} \cite{Kofman:1997yn}. 

Following the formalism reviewed in Section 2, the interaction \eqref{scalar_prod} provides a time dependent mass in \eqref{wsq2} where
$m^2_{eff}(\tau)=g^2 (\varphi - \varphi_0)^2$.  As discussed there, particle production occurs when the adiabatic condition fails\footnote{Within the effective theory, the adiabatic condition can fail either because the particles become massless or simply because the inflaton undergoes non-adiabatic evolution. In the latter case the yield of particles depends on the dynamics of the inflaton and the mass scale of the particles to be produced \cite{Avgoustidis:2012yc}.}, which in this case implies 
$\omega^\prime / \omega^2 \sim m_{eff}^\prime / m_{eff}^2 \gtrsim {\cal O}(1)$. As shown in e.g. \cite{Kofman:2004yc} the production occurs on time scales small compared to the Hubble time so that gravitation effects are negligible and if we denote the time of production as $\tau_0$ the Bogolyubov coefficients 
above for $\tau>\tau_0$ are
\bea
|\alpha_k|^2 &=& 1+\exp\left( -\pi \frac{ k^2}{m^\prime_{eff}(\tau_0)}  \right), \nonumber \\
\left\vert \beta_k \right\vert^2&=& \exp\left( -\pi \frac{ k^2}{m^\prime_{eff}(\tau_0)}  \right), \label{betaeqn}
\eea
where $m^\prime_{eff}(\tau_0) = {g \varphi^\prime_0}$ is the time derivative of the effective mass evaluated at the moment of production and 
we set $a(\tau=\tau_0)=1$.  We also note that the coefficients respect the normalization $|\alpha|^2 - |\beta|^2 =1$ implying that the Bogolyubov transformation is canonical.  The corresponding number density of produced particles is then $n_\chi \sim \int d^3k \; n_k$ with $n_k \sim |\beta_k|^2$ given by \eqref{betaeqn}.  The most interesting case will be when a number of the locations $\varphi=\varphi_0^{(i)}$ occur, as this will lead to a continuous production of gravity waves whereas a single event will lead to an isolated (but perhaps interesting) signature \cite{Senatore:2011sp}. 

The amount of gravitational radiation resulting from both single and multiple events has been examined taking two different approaches.
In \cite{Cook:2011hg} the authors (CS) argued that gravity waves will result from both the production events themselves, as well as from the existence of $\chi$ particles following production -- both sources were found to yield a comparable amount of gravitational radiation.  Whereas in \cite{Senatore:2011sp}, the authors (SSZ) performed estimates adapting the methods of Weinberg \cite{weinberg:1972} to the time dependent case of interest here and found that the production event along with gravitational Bremsstrahlung from the inflaton could result in gravity waves.  Here we will qualitatively argue that the two approaches yield similar results, but for the majority of our calculations we will primarily follow the approach of \cite{Cook:2011hg}.  The key will be that in all instances -- and independent of the computational method -- we find that if particle production for fields coupled directly to the inflaton is to lead to an observable gravity wave signal it presents a tension with existing constraints from Planck on the level of equilateral non-Gaussianity. 

To calculate the gravity waves generated from the presence of on-shell $\chi$ particles following production we can use the result for the Bogolyubov coefficients \eqref{betaeqn} and Green function \eqref{greeny} in the correlator \eqref{result} and we find\footnote{This result agrees with the correlators in both the adiabatic and non-adiabatic cases studied in \cite{Cook:2011hg}.}
\be \label{2ptcs}
\langle h_{ij}(k) \, h_{ij}(k^\prime) \rangle =\frac{ \delta^{(3)}(k+k^\prime)}{2 \pi^5 k^3} \left( \frac{H}{m_p}\right)^4 \left(  \frac{n_\chi }{H^3} \right) F(k\tau_0)
\ee
with $n_\chi \sim (m_{eff}^\prime)^{3/2} = (g \varphi^\prime)^{3/2}$ the number density of produced particles 
and 
\be \label{Ftau}
F(k\tau_0) \simeq \frac{\left[k\tau_0 \cos(k\tau_0) - \sin(k\tau_0) \right]^2}{|k \tau_0|^3} \times \log ^2\left( \frac{n_\chi}{H} \right) \simeq {\cal O} (10 - 100),
\ee
the first term results from the two copies of the Green function \eqref{greeny} in \eqref{result} and peaks around $|k \tau_0| = 2.5$ after which it sharply drops off reflecting both the locality of the production
as well as the fact that only gravity waves produced near the horizon have a chance of contributing significantly to the spectrum\footnote{Causality requires gravity wave production to occur on near or sub-Hubble scales and gravity waves produced on small length scales will undergo significant red-shifting reducing their significance.  Thus, production near the Hubble scale will provide the largest contribution to the tensor spectrum.}.  The second term in \eqref{Ftau} is to be evaluated at the time of production and as we will see is typically at most ${\cal O}(100)$. 

Using the definitions \eqref{tensors} and \eqref{power} the contribution to the tensor power spectrum from production is then
\be \label{newpower}
\Delta_t^2=\Delta^2_{std} \left[ 1+ 4.8 \times 10^{-4}  \left( \frac{H}{m_p}\right)^4 \left(  \frac{n_\chi }{H^3} \right) F(k\tau_0) \right],
\ee
where $\Delta^2_{std}= 2H^2 / (\pi^2 m_p^2)$ is the standard vacuum contribution coming from \eqref{power}.

Before proceeding, let us compare the result \eqref{2ptcs} (and so also \eqref{newpower}) with the estimates found in \cite{Senatore:2011sp}.
From \eqref{2ptcs} we find
\be \label{csresult}
h_{cs}^2 \sim h(k)^2 k^{3} \sim \left( \frac{n_\chi}{H^3} \right) \left(\frac{H}{m_p}\right)^4
\ee
where $h_{cs}$ is the amplitude in real space and $H$ is the Hubble scale during inflation.
Now let us compare this estimate to the one in \cite{Senatore:2011sp}.
There it was found 
\be
h_{ssz}^2 \sim \frac{\rho_{GW}}{\rho_{total}} \sim f \frac{H^3}{E m_p^3},
\ee
where $\rho_{GW} / \rho_{total}$ is the relative energy density in gravity waves, $E$ is the characteristic energy, and $f \sim E n_\chi / \rho_{total}$ is the fraction of waves resulting from the $n_\chi$ density of particles.  Here the frequency of produced waves was taken as $\omega \sim H$ so no red-shifting occurred (as we also assumed in \eqref{csresult}) and plugging in $f$ we find
\be \label{comresult}
h_{ssz}^2 \sim f \frac{H^3}{E m_p^3} \sim \left( \frac{E n_\chi}{H^2 m_p^2} \right) \left( \frac{H^3}{E m_p^3} \right) \sim \left( \frac{n_\chi}{H^3} \right) \left(\frac{H}{m_p}\right)^4 \sim h^2_{cs},
\ee
and so we see the two approaches agree qualitatively.  This result is easily understood -- the production events are independent and so proportional to $n_\chi$ per Hubble volume ($H^3$) and the amplitude of each waves is proportional to $H^2 / m_p^2$ as expected.  

We would now like to see if the new contribution in \eqref{newpower} can be competitive with the vacuum contribution $\Delta^2_{std}$.
We define the difference as 
\be \label{defdP}
\Delta P_t \equiv (\Delta_t^2 - \Delta^2_{std})/\Delta^2_{std}
\ee
so that ultimately we are interested in whether $\Delta P_t \gg 1$ is feasible.
Given \eqref{newpower} we can already see that a large gravity wave signal is difficult to obtain.  In order that the produced particles $n_\chi$ don't ruin inflation we must have at least $n_\chi \ll H^2 m_p^2$.  Using this in \eqref{newpower} we find
\bea
\Delta P_t \simeq10^{-2}  \left( \frac{H}{m_p}\right)^4 \left(  \frac{n_\chi }{H^3} \right) \ll 10^{-2}  \left( \frac{H}{m_p}\right)^2 \left(  \frac{H}{m_\chi} \right),
\eea
where we used that $F(k \tau_0) \lesssim {\cal O}(100)$ and we see that unless the produced scalars remain far lighter than the Hubble scale a competitive signal is simply not possible.
This requires that the gravity waves are produced at the time the field is light ($m_\chi \ll H$) and at the time particle production is occurring.  But this implies that the gravity waves will actually be a scale-dependent feature in the spectrum, which is manifest from the $k\tau_0$ dependence in \eqref{Ftau}.   Instead we are interested in the continuous generation of gravity waves, which suggests that the multiple production case is of more interest.  

Requiring the inflaton to copiously and continuously produce particles, while also providing an adequate number of e-foldings of slow-roll inflation, requires a delicate approach to model building.  However, in models of Trapped Inflation \cite{Green:2009ds} it is precisely this type of balance (and accounting for the backreaction of produced particles) that permits slow-roll inflation in the presence of a steep inflationary potential.  
Denoting the spacing between the particle production events as $\Delta \equiv \varphi_{i+1}-\varphi_i$,
the scalar power spectrum in this model takes the form \cite{Green:2009ds} 
\be
\label{trappedpower}
k^3 P_\zeta  = \frac{g^{7/2} H \varphi^{\prime \, 1/2}}{ \tilde{m} \Delta} \, ,
\ee  
where $\tilde{m}^2 = \frac{7}{2} \frac{g^{5/2}}{\Delta \, (2\pi)^3} \, \varphi^{\prime \, 3/2}$.

In addition, the production events generate non-Gaussianity of the equilateral type, which was estimated in \cite{Green:2009ds} to be
\be \label{trappedng}
f_{\mbox{\tiny NL}}^{\mbox{\tiny equil}} \sim \frac{\tilde{m}^2}{H^2} = 
\frac{7}{2} \frac{g^{5/2}}{\Delta \, (2\pi)^3} \, \frac{\varphi^{\prime \, 3/2}}{H^2} \, .
\ee
Using \eqref{trappedpower} and \eqref{trappedng} we will be able to place constraints on the level of gravity waves resulting from particle production events \cite{Enqvist:2004ey}.

\begin{figure}[t!]
\begin{center}
\includegraphics[scale=0.4]{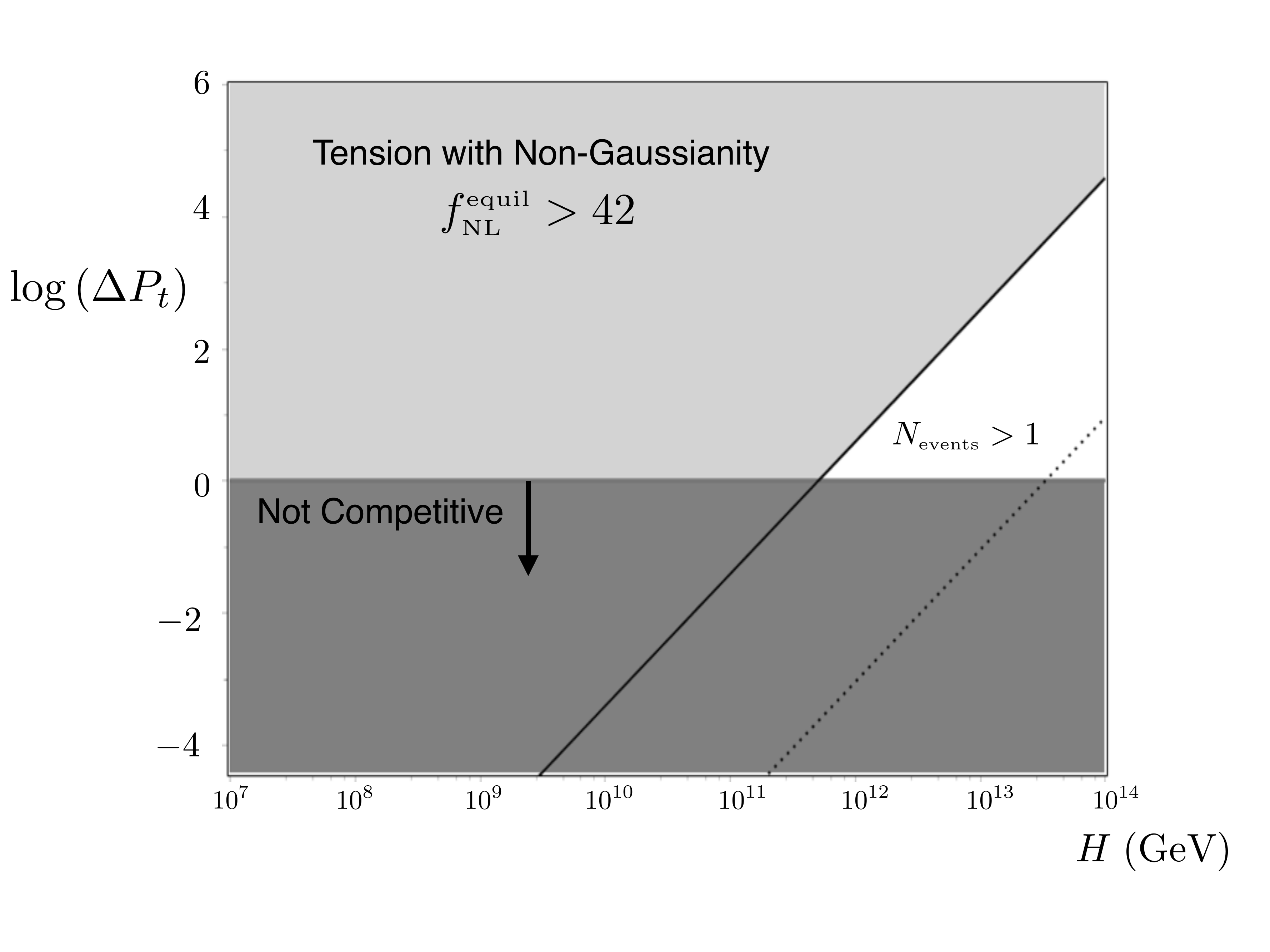}
\end{center}
\caption{Change in the tensor power spectrum due to particle production $\Delta P_t \equiv (\Delta_t^2 - \Delta_{std}^2 )/ \Delta^2_{std}$ with $\Delta^2_{std}$ the contribution from vacuum fluctuations during inflation for both the single ($N_{events}=1$) and multiple production cases ($N_{events}>1$).  The lightly shaded region (gray) represents a tension with the Planck upperbound on equilateral type non-Gaussianity ($|f^{\mbox{\tiny equil}}_{\mbox{\tiny NL}}|<42$), whereas in the dark shaded region particle production is clearly not competitive with the vacuum contribution.  We note that even in cases where the signal is competitive, this does not necessarily imply a dominant contribution.  Thus, we see that in both cases (single or multiple production) non-Gaussianity puts strong constraints on the tensor contribution. The multiple production case is rather insensitive to the coupling, but in both cases we have plotted results for $g=10^{-2}$ and $|f^{\mbox{\tiny equil}}_{\mbox{\tiny NL}}|=42$. \label{trapplt}}
\end{figure}

As before the largest signal will come from gravity waves produced near the Hubble scale  (more precisely near $k\tau_0 =2.5$) as these 
modes will suffer less red-shifting before freeze-out.   In addition, as discussed in \cite{Green:2009ds} the production events are independent 
and so we can simply add the contributions to the tensor spectrum as
\be \label{sum_power}
\Delta P_t^{total} = N_{events} \Delta P_t,
\ee
with $\Delta P_t$ the contribution from a single event 
and the number of events within a Hubble time is roughly $N_{events} = H^{-1} / \Delta t = \varphi^\prime / (H \Delta)$. 
Using \eqref{trappedpower} and \eqref{trappedng} to eliminate $\varphi^\prime$ and $\Delta$ from the tensor spectrum given by \eqref{newpower} and \eqref{sum_power} 
and we find
\be \label{deltap}
\Delta P_t \simeq 3.8 \times \; \left(\frac{N_{events}}{4400}\right)  \left(\frac{H}{10^{12} \; \mbox{GeV}} \right)^4 \left( \frac{g}{0.01}\right)^3 \left( \frac{|f^{\mbox{\tiny equil}}_{\mbox{\tiny NL}}|}{42}\right)^{3/4} \left( \frac{F(k \tau_0) }{ 34}\right),
\ee
where $F(k\tau_0)$ is again given by \eqref{Ftau} and using the fiducial values above and with $|k \tau_0| \simeq 2.5$ we have
\be \label{ftauagain}
F^{1/2}(k\tau_0) = \frac{2}{3} \; \log \left( \frac{n_\chi}{H} \right)\simeq \; \left( 5.8 + \frac{2}{3} \log \left[ \left( \frac{g}{0.01} \right) \left( \frac{|f^{\mbox{\tiny equil}}_{\mbox{\tiny NL}}|}{42} \right)^{1/4} \right] \right),
\ee
where we have again used the constraints \eqref{trappedpower} and \eqref{trappedng}.  
As written, the level of gravity waves seems to depend sensitively on the model parameters, particularly the coupling $g$.  However, again using the constraints and noting that 
the number of events in \eqref{deltap} can also be written as
\be
N_{events}= 2.7 \times 10^{-4} \; \frac{{|f^{equil}_{NL}|}^{3/4}}{g^3} \simeq 4400 \left( \frac{0.01}{g}\right)^3 \left( \frac{|f^{equil}_{NL}|}{42}\right)^{3/4},
\ee
we can reexpress \eqref{deltap} as
\be \label{aresult}
\Delta P_t \simeq 3.8 \times \;   \left(\frac{H}{10^{12} \; \mbox{GeV}} \right)^4  \left( \frac{|f^{\mbox{\tiny equil}}_{\mbox{\tiny NL}}|}{42}\right)^{3/2} \left( \frac{F(k \tau_0) }{34}\right),
\ee
which is only sensitive to the coupling $g$ logarithmically through the dependence in \eqref{ftauagain}.  The strong dependence on the Hubble scale in 
\eqref{aresult} demonstrates the challenge of being consistent with the level of non-Gaussianity while generating a competitive signal.
For the fiducial value $H=10^{12} \gev$ the standard contribution is of the same order of magnitude and reducing to $H=10^{11} \gev$ already rules out particle production as the primary origin.
This can be seen from Figure \ref{trapplt} for both single ($N_{events}=1$) and multiple production ($N_{events}>1$) cases. We emphasize that our result \eqref{aresult} shows the {\em tension with non-Gaussianity constraints without even invoking model building constraints within Trapped Inflation}.  Moreover, as bounds on non-Gaussanity improve this will strengthen confidence in vacuum fluctuations as the origin of the primordial tensor spectrum.

\subsection{Pseudo-scalar Inflaton and Vector Production}\label{PSI}
A promising candidate for large field inflation is when the inflaton is realized as a pseudo-Nambu-Goldstone boson (PNGB)
associated with the breaking of a global symmetry at some high scale \cite{Freese:1990rb,Adams:1992bn}.  
As discussed above, the UV sensitivity of large field inflation emphasizes the importance of realizing
models of inflation within a high energy framework.  Within string theory constructions, PNGB's naturally arise from the compactification of higher dimensional gauge fields to four dimensions \cite{Svrcek:2006yi}.
Often the higher dimensional gauge invariance of these fields leads to an an approximate shift symmetry $\varphi \, \rightarrow \, \varphi \, + \, \, {\rm const.}$ in the $4D$ low-energy effective theory.
This shift symmetry can be lifted by a number of effects (both tree-level and non-perturbative) that depend on the details of the compactification \cite{Svrcek:2006yi}.  
One promising class of PNGB inflation models are those arising in models of axion monodromy (see \cite{McAllister:2014mpa} and references within).  
In these models a large field range for the inflaton is achieved as branes wrapping the same extra dimensions as the gauge fields lift the shift symmetry in a controlled way leading to relevant terms in the inflaton potential but with naturally suppressed coefficients.  Other non-perturbative effects can contribute (such as gauge and brane instantons), but in particular constructions its possible to arrange for these effects to be parametrically small yielding a viable inflation model.  That is, these string based models realize the idea of natural inflation proposed in \cite{Freese:1990rb,Adams:1992bn} in a technically natural way.

In addition to the inflaton sector, it is natural for PNGB inflatons to couple to $4D$ gauge fields.  In fact, for any PNGB of inflation it is natural to consider interactions of the form
\be \label{gauge_prod}
\mathcal{L}_{int} \, = \, - \frac{1}{4 f} \varphi F^{\mu \nu} \tilde{F}_{\mu \nu} \,\,,
\ee
where $F_{\mu \nu} = \partial_\mu A_\nu - \partial_\nu A_\mu$ and $\tilde{F}=(1/2) \epsilon_{\mu \nu \lambda \sigma} F^{\lambda \sigma}$ is the dual field strength\footnote{Here we are interested in the case that the inflaton is a pseudo-scalar and so we do not consider couplings of the type $\sim f(\varphi) F_{\mu \nu} F^{\mu \nu}$ -- see e.g. \cite{Barnaby:2012tk} for these scalar inflaton models.  However, we expect the bounds on the level of gravity waves found in this section and the previous section to be representative of any inflaton coupled to the fields being produced.}. The `axion' decay constant is denoted by $f$. For an exactly constant field $\varphi$ this term is a total derivative (i.e. topological), and so it does not enter the equations of motion.  However, during inflation the inflaton slowly evolves and this interaction leads to an effective tachyonic-like mass term for the gauge field -- thus, particle production is possible. We are interested in whether such a term can lead to significant gravity wave production, while evading the bounds established for the scalar production case in the last section.

Just as in the scalar case of the previous section, the interaction \eqref{gauge_prod} can lead to particle production of gauge fields $\delta A$ when adiabaticity is violated\footnote{See \cite{Barnaby:2011qe} for a review.}. 
Analogous to \eqref{wsq1} we have
\be \label{gaugewsq1}
{A}^{\prime \prime}_{\pm} \, + \, \omega^2(\tau) A_{\pm} \, = \, 0 \,\, ,
\ee
where $\pm$ denotes the transverse polarization. 
The time dependent frequency of the field is
\be \label{somega} 
\omega^2(\tau) = k^2 \mp m^2(\tau,k) \,\,,
\ee
where the tachyonic-like mass term depends on the wave number as
\bea \label{tmass}
m^2(\tau, k)&=& \frac{k \dot{\varphi}}{H|\tau|f}, \nonumber \\
&=& \frac{1}{2 \pi \sqrt{\Delta_s^2}} \left(\frac{kH}{f |\tau|}\right) \simeq 3.2 \times 10^{3} \; k^2 \left(\frac{H}{f}\right)\left(\frac{1}{k|\tau|}\right)
\eea
where in the last line we have used the normalization of the scalar power spectrum $\Delta_s^2 = 2.4\times 10^{-9}$ to eliminate\footnote{We note that for comparison with the results in \cite{Barnaby:2010vf,Barnaby:2011vw}, here we have used the COBE normalization to simplify the $\xi\equiv\dot{\varphi}/{(2Hf)}$ parameter of that paper where one would find $\xi=(H/f)/(4 \pi \sqrt{\Delta_s^2})$.} $\dot{\varphi}$.  Assuming $\dot{\varphi}>0$ without loss of generality, it is easy to see from \eqref{gaugewsq1},\eqref{somega} and \eqref{tmass} that only positive helicity gauge modes $A_+$ are amplified while $A_-$ modes stay in the vacuum. For $\dot{\varphi}/{(2Hf)} \simeq constant$ the amplification of the positive helicity modes is given by \cite{Barnaby:2010vf,Barnaby:2011vw}
\be \label{Asimpleeq}
A_{+} \,\, \sim \,\, e^{(H/f)/(4\sqrt{\Delta_s^2})}  \,\,\,\;\;\; {\rm for} \,\,\,\;\;\; k|\tau| \, \lsim \, (H/f)/(2 \pi \sqrt{\Delta_s^2}) \,\,,
\ee
where the exponential dependence here is consistent with the particle production coming from a tachyonic instability. The essential physics is that gauge field modes with $k|\tau| \, \lsim \, (H/f)/(2 \pi \sqrt{\Delta_s^2})$ will become violently excited by the interaction as the tachyonic mass term \eqref{tmass} becomes significant in \eqref{somega} in this regime. On the other hand, this growth saturates deep in the IR $k|\tau|\to 0$ causing the ``physical" $\vec{E}$ and $\vec{B}$ fields to decay sufficiently far outside the horizon\footnote{We thank Lorenzo Sorbo for discussions related to this point.}.  This can be seen by the scaling of the physical fields with the expansion, i.e. $\vec{B}=(\vec{\nabla} \times \vec{A})/a^2$ and $\vec{E}= - \vec{A}^\prime / a^2$.  
 Whereas the gravity waves that result from this process will not decay outside the Hubble radius, but instead become 'frozen-in' with this process over time leading to a late time stochastic background of gravity waves. The question is whether this source is competitive with that of the quasi-dS vacuum fluctuations. To answer this question, first we need to take into account the constraints arising on non-Gaussianities and back-reaction produced from the interaction \eqref{gauge_prod}.

As shown in \cite{Barnaby:2010vf,Barnaby:2011vw}, the NG contribution to cosmological correlators in this model arises due to the inverse decay processes: $\delta A  \delta A \to \delta\varphi$ associated with the interaction \eqref{gauge_prod}. This new source of inflaton fluctuations leads to curvature perturbations $\zeta \sim -(H/\dot{\varphi})\delta\varphi$ and non-Gaussianity of the equilateral type. 
Following \cite{Barnaby:2011vw} the parameter range of interest implies that the correction to the scalar power spectrum is negligible and the resulting tensor spectrum 
in our notation is
\be
\Delta P_t \simeq (\Delta_s^2)^3 \left( \frac{H}{m_p} \right)^2 \left( \frac{f}{H} \right)^6 \exp\left( \frac{H}{f\sqrt{\Delta_s^2}} \right),
\ee
Whereas, the equilateral type NG is  \cite{Barnaby:2011vw}
\be
|f^{\mbox{\tiny equil}}_{\mbox{\tiny NL}}| \simeq 2.2 \times 10^{3} \; \left(\Delta_s^2\right)^{11/2} \left( \frac{f}{H}\right)^9 \exp\left( \frac{3H}{2f\sqrt{\Delta_s^2}} \right)
\ee
Thus, combining these results and utilizing the Planck result $|f^{\mbox{\tiny equil}}_{\mbox{\tiny NL}}|<42$ we find a bound on the tensor spectrum
\bea
\Delta P_t &\lesssim& 5.5 \times 10^3 \; {|f^{\mbox{\tiny equil}}_{\mbox{\tiny NL}}| }^{2/3} \left(\frac{H}{m_p} \right)^2, \nonumber \\
&\lesssim& 1.2 \times 10^{-8} \; \left(\frac{|f^{\mbox{\tiny equil}}_{\mbox{\tiny NL}}| }{42} \right)^{2/3} \left(\frac{H}{10^{12} \; \mbox{GeV}} \right)^2, 
\eea
and we see it is difficult for this model to account for the gravity wave spectrum while allowing for a low-scale inflation model.

\subsection{Summary of Direct Coupling Case}
In this section, we have considered particle production resulting from a direct coupling of fields to the inflaton and the resulting production of 
gravitational waves.  In particular, we were interested in whether the contribution to the tensor power spectrum from production events
could be the leading contribution, since this would imply that an observation of primordial tensors does not necessarily imply the scale at which inflation took place.

However, in both the scalar and gauge field cases we have seen that existing constraints on non-Gaussianity from Planck lead to a tension 
for model building if the produced tensor signal is to be competitive with the quasi-dS source.  Moreover, for gauge field production additional constraints from back-reaction make it 
very difficult to see how such a source could lead to an alternative origin of gravity waves for any range of the parameters.  For scalar production we saw that multiple events can improve the situation, however we are left with a small region of the parameter space near the highest possible inflationary scales.  Therefore, even in this special region, if a substantial signal resulted it would still provide us with information on the (high) scale at which inflation took place.

Given the direct coupling of the produced fields to the inflaton in these models a strong level of constraint from non-Gaussianity bounds was anticipated --
our results have quantified this.  However, we have also argued that the types of couplings and interactions above are generic expectations from the UV perspective, e.g. in the context of inflationary model building within string theory.  One may have asked if such interactions are around, why have we not seen their gravitational signatures?
Our results imply that these fields can exist and be produced without leading to a large tensor contribution.

In the next section, we consider gravity wave and particle production in models that contain fields which are only gravitationally coupled to the inflaton.

\section{Particle Production Mechanisms with Gravitational Coupling \label{section4}}
In this section we would like to explore whether particle production in a hidden sector which is only gravitationally coupled to the inflaton can lead to a competitive alternative for generating a primordial tensor spectrum.  As in the previous section, we will again utilize constraints on the back-reaction and on the level of non-Gaussainity -- the latter anticipated to be less stringent since the fields are only gravitationally coupled.

The system is described by the following action \cite{Barnaby:2012xt,Cook:2013xea}
\be \label{S}
S= \int d^4x \sqrt{-g}~\left[ \frac{1}{2} m_p^2 R - \frac{1}{2} (\partial \vp)^2 - V(\vp)  + \mathcal{L}_{hidden}[\partial_{\mu}\chi,\chi ,F]\right],
\ee
where the field $\vp$ is the inflaton and we assume $V(\vp)$ can support inflation.
$\mathcal{L}_{hidden}$ consists of a pseudoscalar field $\chi$ with potential $U(\chi)$ during inflation. It is coupled gravitationally to the inflaton and $U(1)$ gauge field $A_{\mu}$ through an axionic coupling,
\be\label{Lhidden}
\mathcal{L}_{hidden}= -\frac{1}{2} (\partial \chi)^2 - U(\chi)-\fr{1}{4}F_{\mu\nu}F^{\mu\nu}-\fr{\chi}{4f}F_{\mu\nu}\tilde{F}^{\mu\nu}.
\ee 
The gauge field production is similar to the case in Section 3, except this time the $\chi$ field is responsible for the amplification of the gauge field fluctuations.
The tachyonic mass term responsible for amplification is now $m^2(k,\tau)=  k^2  \dot{\chi} /(k|\tau| Hf)$ and modes grow as
\be \label{Athing}
A_{+} \, \sim \, e^{\pi \sqrt{\frac{\epsilon_{\chi}}{2}} \left(\frac{m_p}{f} \right) }  ~~~~~{\rm for}  ~~~~~
k|\tau|\lesssim \sqrt{2 \epsilon_{\chi}} \left(\frac{m_p}{f}\right) . 
\ee
The parameter $\epsilon_{\chi}$ is given by $\epsilon_{\chi} = \dot{{\chi}}^2/(2H^2 m^2_p)$. 
\vspace{0.1in}

\noindent
Successful model building requires the following conditions:
\bi
\vspace{-0.1in}
\item
The field $\chi$ is to be a spectator field implying that 
\be \label{assm}
U(\chi) \ll V(\varphi), ~~ \dot{\chi}^2 \ll \dot{\vp}^2.
\ee
\item The energy density of the produced gauge fields must be sub-dominant to the kinetic energy of $\chi$ and this energy should not back-react on the background evolution of $\chi$. 
It turns out that the former is a stronger condition than latter \cite{Barnaby:2012xt,Mukohyama:2014gba}. Therefore we require,
\be \label{br1}
\frac{1}{2}\dot{\chi}^2 \gg \frac{1}{2} \langle\vec{E}^2+\vec{B}^2\rangle,~~~~~ 
\ee
where from the earlier definitions for the fields and using the mode functions one can show that this gives \cite{Barnaby:2012xt,Mukohyama:2014gba}
\be \label{big_back}
\dot{\chi}^2 \gg 2.2 \times 10^{-3} \; H^4 \left( \frac{Hf}{\dot{\chi}}\right)^3 \exp\left(\frac{\pi \dot{\chi}}{Hf} \right)
\ee

\item Non-Gaussianity constraints from Planck imply $|f^{\mbox{\tiny equil}}_{\mbox{\tiny NL}}|< 42$.
\ei
Contrary to the pseudo-scalar inflation case, the inverse decay effects ($\delta A \delta A\to \delta\chi$) associated with the last term in \eqref{Lhidden} do not necessarily produce strong NG correlations.
Moreover, as shown in \cite{Barnaby:2012xt,Cook:2013xea,Mukohyama:2014gba} the scalar power spectrum $P_\zeta$ gets a negligible contribution from the gauge fields. 
On the other hand, gravity waves sourced by vector fields $A_\mu$ can dominate over vacuum ones and hence can contribute significantly to the tensor power spectrum $P_t$.
The change in the tensor power spectrum is given by \cite{Barnaby:2012xt,Cook:2013xea,Mukohyama:2014gba}
\be
\Delta P_t \simeq 2.8 \times 10^{-5} \; \left( \frac{H}{m_p} \right)^2 \left( \frac{H f}{\dot{\chi}} \right)^6 \exp\left( \frac{2 \pi \dot{\chi}}{Hf} \right),
\ee
and the NG is 
\be
f_{NL} \simeq 2.5 \times 10^5 \; \left( \frac{H}{m_p}\right)^6 \left( \frac{ H f}{\dot{\chi} } \right)^9 \exp \left({\frac{3\pi \dot{\chi} }{ Hf}}\right).
\ee
Combining these we find the constraint
\be \label{T1}
\Delta P_t \lesssim ~ 4.8 \times 10^{5} \; \left(\fr{|f^{\mbox{\tiny equil}}_{\mbox{\tiny NL}}|}{42}\right)^{2/3}\left(\fr{10^{12}~ {\rm GeV}}{H}\right)^2,
\ee
which demonstrates that the tensor signal can be quite large depending on the inflationary scale.

\begin{figure}[t!]
\begin{center}
\includegraphics[scale=0.4]{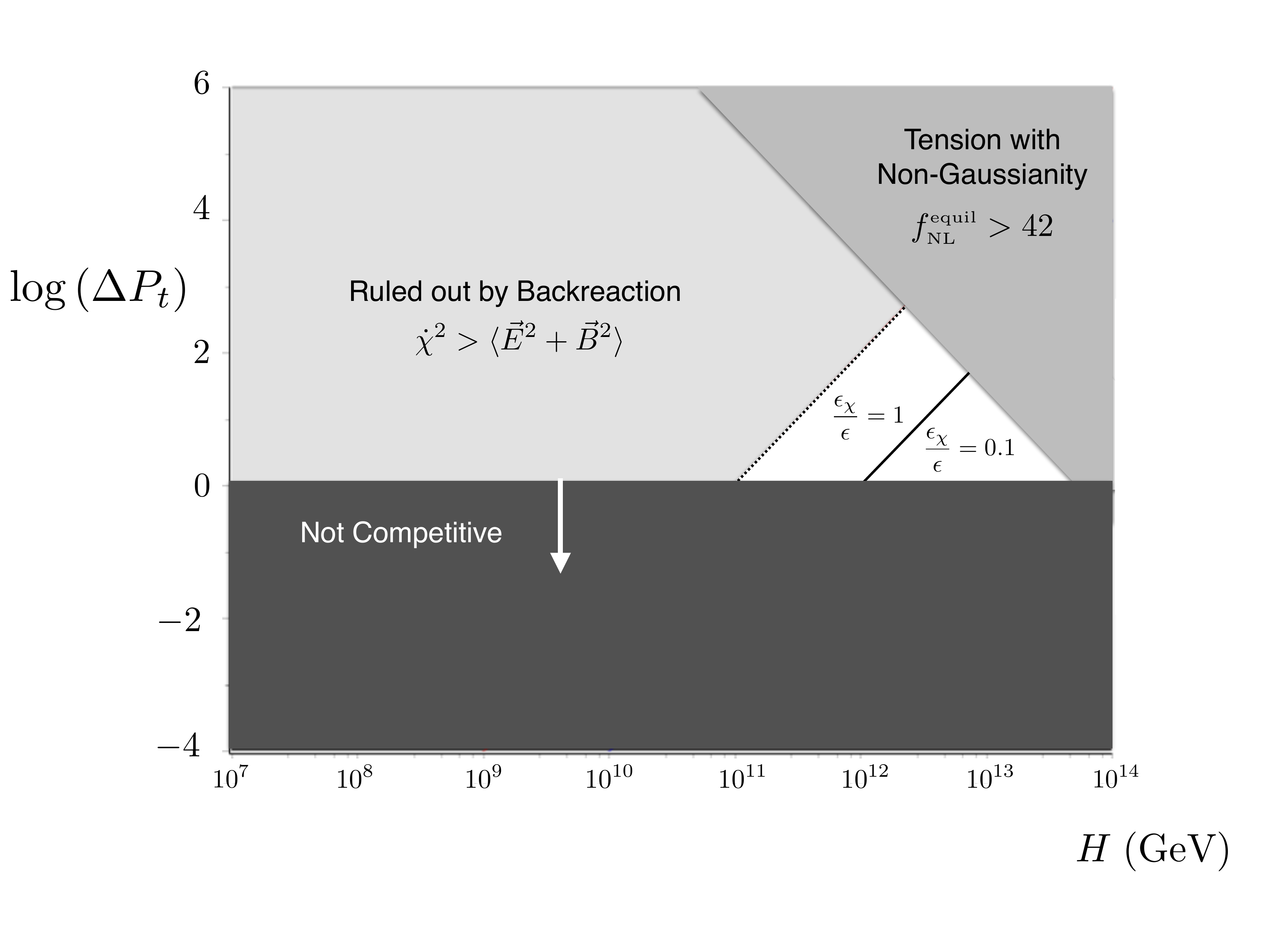}
\end{center}
\caption{\label{figure3} Change in the tensor power spectrum $\Delta P_t \equiv (\Delta_t^2 - \Delta_{std}^2 )/ \Delta^2_{std}$ 
due to (gravitationally coupled) gauge field production as discussed in the text, with
$\Delta^2_{std}$ the contribution from vacuum fluctuations during inflation. The medium gray region represents a tension with the Planck upperbound on equilateral type non-Gaussianity ($|f^{\mbox{\tiny equil}}_{\mbox{\tiny NL}}|<42$), whereas in the darkest shaded region  gauge field production is not competitive with the vacuum contribution.  The light gray region corresponds to the constraint coming from the back reaction of the produced gauge fields on the spectator scalar $\chi$.  As plotted, the above graph is actually conservative as the real constraint requires the kinetic energy to be {\em much} greater than the gauge field energy $\dot{\chi}^2 \gg \vec{E}^2 + \langle \vec{B}^2 \rangle$. Given this caveat, the two white regions represent the available parameter space for the choices $\epsilon_\chi = \epsilon$ and the more realistic value $\epsilon_\chi < 0.1 \, \epsilon$ ($\epsilon_\chi < \epsilon<1$ is required for $\chi$ to remain a spectator field.).} 
\end{figure}

However, we have not yet used the back reaction constraint \eqref{big_back}.
This turns out to be a much more stringent constraint compared to NG and we find
\bea \label{T2}
\Delta P_t &\ll& 5.7 \times 10^{14} \left( \frac{\epsilon_\chi}{\epsilon} \right)^2 \left(\frac{H}{m_p} \right)^2 \nonumber \\
&\ll& ~ 1.0 \; \left(\fr{\epsilon_\chi/\epsilon}{0.1}\right)^2~\left(\fr{H}{10^{12}~{\rm GeV}}\right)^2,
\eea
where $\epsilon$ is the usual slow-roll parameter of inflaton and we took a fiducial value for the ratio $\epsilon_\chi /\epsilon$ that is implied by the condition \eqref{assm}. In Figure \ref{figure3}, we summarize the constraints obtained from \eqref{T1} and \eqref{T2}. These constraints can also be used to restrict the axion decay constant $f$. The requirement of generating a significant tensor signal implies that the argument of the exponential in \eqref{Athing} must satisfy $ \sqrt{\frac{\epsilon_{\chi}}{2}} \left(\frac{m_p}{f} \right)  \gtrsim 3.5$. Using this along with the constraint from \eqref{assm} we have
\be \label{fconstrain1}
\fr{f}{m_p}\ll 1.8 \times 10^{-2} \left(\fr{\epsilon}{0.008}\right)^{1/2},
\ee
where we have chosen a fiducial value for the slow-roll parameter corresponding to a quadratic potential with $N=60$ e-foldings. On the other hand, \eqref{br1} implies $f/m_p\gg 8.5 \times 10^{-4} (\epsilon/0.008)^{1/2}$ and so we have
\be \label{fconstrain2}
8.5 \times 10^{-4} \left(\fr{\epsilon}{0.008}\right)^{1/2}\ll \fr{f}{m_p}\ll 1.8 \times 10^{-2} \left(\fr{\epsilon}{0.008}\right)^{1/2}.
\ee
Thus, we find that this model provides a competitive source of gravity waves for a narrow region of the parameter space.  
We now turn to the question of whether such a model is UV completable. Studies of other UV contexts such as braneworld inflation have recently been undertaken in \cite{Neupane:2014vwa}.

\section{Towards a UV Completion and the Resulting Constraints from String Theory \label{UVchallenges}}

We have seen that the most promising case for observable particle production arises from the gravitationally coupled case.
In this section we want to consider the possibility of UV completing such a model and any additional constraints on model building that this might imply. In the next few subsections, we gather the tools that will be required for our analysis.

\subsection{Axions in Type IIB String Theory}

As our starting point we will focus on axions arising from compactifications of Type IIB string theory.
To see how axions arise in the theory, we consider the dimensional reduction of the theory to four dimensions by starting from the 
$10$D action in the string frame given by \cite{Polchinski:1998rr}
\be \label{act1}
S_{10}^{IIB}=\frac{1}{(2\pi)^7\alpha^{\prime \; 4}}\int d^{10}x \sqrt{-G} \left[  \frac{1}{g_s^{2}} \left(R[G] - \frac{1}{2} |H_3|^2 \right) + \frac{1}{2} |F_3|^2\right] + \ldots
\ee
where $G_{MN}$ is the ten dimensional string frame metric and $H_3=dB_2$ and $F_3=dC_2$ are the NS-NS and RR three-form fluxes, respectively, with $B_2$ and $C_2$ the corresponding gauge potentials and 
$1/(2 \pi \alpha^\prime)$ is the string tension. The model independent axion $C_0$ and dilaton are combined as the axio-dilaton $\tau =C_0 + i/g_s$ where $g_s = \exp(\phi_0)$ is the string coupling and we will take $C_0$ to be fixed and instead concentrate on the model dependent axions arising from the compactification of the form fields.  The additional terms represented by dots will be discussed in more detail below and include higher form fields such as $C_4$ (we will concentrate on $C_2$ axions for now).

The zero modes of $B_2$ and $C_2$ are independent of the co-ordinates of the compact dimensions and can be integrated over chosen two-cycles of the internal geometry giving rise to axions in the four dimensional theory.  To make this explicit, consider compactifying on a Calabi-Yau 3-fold ($CY_3$) and for the form field $C_2$ we make the ansatz \cite{Svrcek:2006yi}
\be \label{axionci}
C_2= \frac{1}{2\pi} c_I(x)\omega^I,
\ee
where the $c_I(x)$ are only functions of the four non-compact space-time dimensions and $I$ labels the two-cycle.  We have introduced the basis forms $\omega^I$ to describe the internal geometry and they obey the normalization condition $\int_{\Sigma_I} \omega^J = (2\pi)^2 \alpha^\prime \delta_I^J$ with the two cycles $\Sigma_I$ giving a basis of the dual homology $H_2(X,\mathbb{Z})$.  The normalization factors of $2\pi$ are chosen for later convenience. Making a similar ansatz for $B_2$ and using this in \eqref{act1} we have
\be
S=\int \frac{d^{10}x }{(2\pi)^7\alpha^{\prime \; 4}} \sqrt{- G} \left[ g_s^{-2}\left(R[G]  - \frac{1}{48\pi^2} G^{nm}G^{lp} \partial_\mu b_{I} \partial^\mu b_{J} \omega^I_{nl} \omega^J_{mp} 
 \right) - \frac{1}{48 \pi^2} G^{nm}G^{lp} \partial_\mu c_{I} \partial^\mu c_{J} \omega^I_{nl} \omega^J_{mp}  \right], 
\ee
where Greek indices run over the four non-compact dimensions and lower-case latin indices denote the compact dimensions.  
 At the classical level, the gauge invariance of the higher dimensional gauge potential implies that the axions can only be derivatively coupled and so we have a shift-symmetric pseudo-scalar in the low energy theory.
This symmetry can be broken in a number of ways, which we will discuss shortly.  
Denoting both types of axions as $a_I$, upon dimensional reduction we find
\be
S_4 = \int d^4x \sqrt{-g} \left( \frac{m_p^2}{2} R[g] - \frac{1}{2} \gamma^{IJ} \partial_\mu a_I \partial^\mu a_J  \right)
\ee
where $g$ is the $4D$ Einstein frame metric and the $4D$ reduced Planck mass is
\be \label{dimreducedplanck}
m_p^2 = \frac{2 {\cal V}}{(2\pi)^7 g_s^2 \alpha^\prime},
\ee
with ${\cal V} =V_6/\alpha^{\prime \; 3}$ the string frame volume of $CY_3$. 
Another common convention is to instead work with the Einstein frame volume.
The $10$D string frame is related to the Einstein frame by the 
Weyl rescaling $ G_{MN}^{string} = \exp(\phi/2) G_{MN}^{Einstein}$ and working in units of the string length  
$l_s=2\pi \sqrt{\alpha^\prime}$ the two volumes are related by ${\cal V}_E= (2\pi)^6 {\cal V} /g_s^{3/2}$.

For the RR axion the $\gamma^{IJ}$ provide the axion decay constants and depend on the internal geometry as \cite{Svrcek:2006yi}
\be \label{ohmyomega}
\gamma^{IJ}_{\mbox{\tiny RR}}=\frac{1}{6(2\pi)^9 \alpha^{\prime \; 4}} \int \omega_I \wedge \star \, \omega_J,
\ee
whereas for the NS axion one gets the same result multiplied by an extra factor of $g_s^{-2}$.  
However, for the remainder of this paper we will restrict our attention to the RR axion, since 
(as shown in \cite{McAllister:2008hb}) the NS axions will suffer an $\eta$ problem when moduli stabilization proceeds through non-perturbative effects\footnote{This is because the NS axion appears explicitly in the Kahler potential and leads to a Hubble scale mass for the field in the models we will consider.  For recent progress on this issue in a different class of models we refer the reader to \cite{McAllister:2014mpa}.}.  Alternatively, one could work in perturbative (Large Volume) stabilization scenarios \cite{Cicoli:2012sz}.

Once the specifics of the internal geometry are known one can calculate the $\gamma^{IJ}$ to find the corresponding axion decay constants. This is a non-trivial task, which requires a full specification of the compactification geometry in a Calabi-Yau manifold. Mostly, we will be interested in order-of-magnitude estimates of this quantity. This is the project taken up in \cite{Svrcek:2006yi}, where the quantity $\gamma^{IJ}$ was calculated in a variety of string models assuming compactifications sufficiently symmetric to be amenable to estimates.

\subsection{Orientifold Compactification Data and Axion Decay Constant} \label{orintdataax}

In order to proceed with concrete estimates of the axion decay constant and the axion potential, it is best to locate oneself within a orientifold compactification in which these quantities can be given in terms of the compactification data.

We will consider an $N=2$ IIB compactification on $CY_3$, which has a moduli space $\CM_h \times \CM_v$ of exactly flat directions. Here, $\CM_h$ denotes the hypermultiplet moduli space while $\CM_v$ is the vector multiplet moduli space. $\CM_h$ is a quaternionic manifold whereas $\CM_v$ is a special K\"ahler manifold. The dilaton field is a  hypermultiplet component, implying that the geometry of $\CM_h$ receives both $\alpha'$ and $g_s$ corrections. The geometry of $\CM_v$, on the other hand, is exact at tree level in both $\alpha'$ and $g_s$. The hypermultiplet moduli space $\CM_h$ contains a subspace $\CM_h^0$ which is parameterized by vacuum expectation values of NS-NS fields, with the RR moduli being set to zero. At string tree level $\CM_h^0$ has a special Kahler structure that receives nonperturbative $\alpha'$ corrections which can be exactly summed using 
mirror symmetry. 

From this $N=2$ compactification, we can construct a $N=1$ theory by gauging a discrete symmetry of
the form $(-1)^{\epsilon F_L}\Omega\sigma$ where $\Omega$ denotes world-sheet parity,
$F_L$ is left-moving fermion number and $\epsilon$ takes values $0,1$ depending on the model. Note that $\sigma: X \to X$ is a holomorphic involution of the Calabi-Yau manifold $X$ which preserves the holomorphic three-form $\Omega_X$ up to sign $\sigma^* \Omega_X = (-1)^\epsilon \Omega_X.$ For the purposes of this paper, we will take $\epsilon=1$, which corresponds to theories with O3/O7 planes. 

The analysis of \cite{Grimm:2004uq} tells us that the massless spectrum of $N=1$ orientifold compactifications is naturally organized in vector and chiral multiplets. For orientifolds with O3/O7 planes, there are $h^{2,1}_-$ chiral multiplets which correspond to invariant complex structure deformations of $X$, $h^{1,1}_+$ chiral multiplets that correspond to invariant complexified K\"ahler deformations of $X$, and $h^{1,1}_-$ chiral multiplets that parameterize
the expectation values of the two-form fields $B_2$ and $C_2$. Moreover, there is a dilaton-axion modulus $\tau$. The real K\"ahler deformations 
of $X$ pair up with expectation values of the four-form field $C_4$, giving rise to the $h^{1,1}_+$ complexified K\"ahler moduli. 

The moduli space of the $N=1$ theory is a K\"ahler manifold. In the limit of small string coupling and large compactification radius the moduli 
space is a direct product of the complex structure moduli, complexified K\"ahler moduli and a dilaton-axion factor. The K\"ahler geometry of the moduli is determined in this regime by KK reduction of ten dimensional supergravity \cite{Grimm:2004uq}. For more general values of parameters, however, the geometry receives both $\alpha'$ and $g_s$ corrections which does not preserve the direct product structure. In particular, significant $\alpha'$ corrections are expected in nongeometric regions of the K\"ahler moduli space such as the Landau-Ginzburg phase \cite{Witten:1993yc,Diaconescu:2006id}. 

In this paper, we will stay in the geometric phase. The moduli space of the theory has a direct product structure 
\be\label{eq:modspace}
\CM \times \CK
\ee
where $\CM$  and $\CK$  are the complex structure and K\"ahler moduli space respectively of the IIB orientifold 
$(X,\sigma)$. 

To further discuss the geometry of $\CK$, it is necessary to introduce certain geometric data. The K\"ahler potential is given as $K_{\CK} = -2 \ln{\mathcal{V}_E}$ in terms of the  dimensionless volume $\mathcal{V}_E$ in the Einstein frame. The Kahler form is given by $J=v_\alpha \omega^\alpha$ and the volume by ${\cal V}_E=\frac{1}{6} \frac{\int J \wedge J \wedge J}{(2 \pi \sqrt{\alpha^\prime})^6} =  \frac{1}{6} v^I v^J v^K c_{IJK}$, where the triple intersection numbers are given by
\be
c_{IJK} = \frac{1}{(2 \pi \sqrt{\alpha^\prime})^6} \int \omega_I \wedge \omega_J \wedge \omega_K \,\,.
\ee
The $\omega_I$ are the basis of the cohomology of $H^2(X,\mathbb{Z})$ with normalization $\int_{\Sigma_I} \omega_J = (2 \pi)^2 \alpha^\prime \delta^I_J$. The two-cycle volumes $v^{i}$ are functions of the appropriately defined K\"ahler coordinates $T_{i} = (3i/2) \rho_i + (3/4)c_{ijk} v^j v^k - (3/2)\zeta_i$ and $G^{i} = (1/2\pi)(c^i - i(b^i/g_s))$. Here, $\zeta_i = -(i/2(\tau - \overline{\tau}))c_{ijk} G^j (G - \overline{G})^{k}$ and the $c_i$ have been defined in \eqref{axionci}, with similar expressions for $b_i$.

The axion decay constant can now be extracted in terms of the orientifold data by noticing that $\gamma_{IJ}$ given in \eqref{ohmyomega} is the K\"ahler metric $K_{G \overline{G}}$ along the axion direction. For an axion wrapped on the two-cycle $\Sigma$ we have \cite{Grimm:2004uq,Flauger:2009ab}
\be
 \frac{1}{l_s^{6}} \int \omega_I \wedge \star \, \omega_J = \frac{2}{3}c_{\alpha \sss \sss} v^\alpha
\ee
where $v^\alpha$ is the dimensionless volume of the two-cycle in string units, the $\alpha$ index runs over the number of two-cycles surviving the orientifold projection, $\Sigma$ is the two-cycle wrapped by the axion.
Using this and \eqref{ohmyomega} 
the axion decay constants are then
\be \label{decay_constants}
\left( \frac{f_\sss}{m_p}\right)^2 =\frac{g_s}{8 \pi^2} \left( \frac{c_{\alpha \sss \sss} v^\alpha }{{\cal V}_E}\right)
\ee

As a simple example, if we consider an internal geometry that is highly symmetric with all two-cycles of equal size $L l_s$
then using \eqref{dimreducedplanck} we have $(f/m_p)^2 \sim g_s {\cal V}_E^{-2/3}$.  Thus, we see that requiring the string theory completion of the axion model explicitly connects the
compactification scale ${\cal V}_E$, the string coupling $g_s$, and the axion decay constant to the Planck scale.  As we will see, theoretical consistency will lead to requirements such as ${\cal V}_E >1$ and $g_s < 1$ (for validity of the geometric regime) leading to additional constraints on model building. 

\subsection{Stable Five-brane-Anti-brane Systems and Axion Potentials}

Given the axion decay constants \eqref{decay_constants}, we now turn to the question of their potential energy. Classically the axions descending from the compactification enjoy a shift symmetry, however there are a number of ways the symmetry can be broken. 

Crucial to our construction will be the presence of either $D5$ branes or $NS5$ branes in the geometry. In this subsection, we discuss various aspects of such geometric constructions.

The D-brane configuration will consist of a 5-brane wrapping a holomorphic curve $\Sigma$ and an anti-5-brane wrapping the image curve $\Sigma^{\prime}$ under the orientifold projection. We will take $\Sigma$ and $\Sigma^{\prime}$ to be rigid cycles that do not intersect each other. Under the orientifold action, the modulus $T_+$ of the even combination $\Sigma_+$ and the modulus $G_-$ of the odd combination $\Sigma_-$ are projected in. The sizes of the cycles are equal in the covering space: $v_{\Sigma} = v_{\Sigma^{\prime}} = \frac{1}{2}v_+$, while the odd volume modulus $v_-$ is projected out. By abuse of notation, we will continue to use $v_{\Sigma}$ to denote the even volume modulus.

In general, one has to be careful about open string fields in the brane / anti-brane sector which may destabilize the system. In flat space one would expect the system to decay and give a supersymmetric configuration of space-filling $D3$ branes. On a $CY_3$, the curves $\Sigma$ and $\Sigma^{\prime}$ can be chosen to be rigid, meaning that the corresponding wrapped branes have no moduli. For branes that are sufficiently far apart, the open string spectrum is not expected to contain tachyons. The attractive force will be weak, resulting in a metastable state which can only decay through tunnelling effects \footnote{There is an added layer of complication for branes with magnetic fluxes. There is a tachyonic contribution to the mass of the lightest open string modes that is proportional to the supersymmetry breaking parameter, which is given by the relative phase between the central charges of the $D5$ and the induced $D3$ \cite{Diaconescu:2006nk}. However, this tachyonic contribution is usually small. We will explore these issues soon for the brane construction of the $\chi$ sector.}.

The situation is best studied by taking a simple potential for the system. We will be interested in the case where $\Sigma$ and $\Sigma^{\prime}$ are part of a one parameter family of holomorphic curves $\mathcal{E}$. The effective dynamics of the brane system can be described by a single chiral superfield $\zeta$, which corresponds to normal deformations of the brane wrapping $\Sigma$. This can also be identified as the normal deformations of the anti-brane wrapping $\Sigma^{\prime}$. The effective dynamics of the system can be described by a potential of the form
\be \label{braneantibranepot}
V(r) = m (r - r_0)^2 \, + \, c \ln{\left(\frac{r}{r_0}\right)} \,\,,
\ee
where $r$ is the distance between the two branes. The first term is a quadratic mass term corresponding to normal deformations of the brane in the ambient $CY_3$. The second term is a typical two dimensional brane / anti-brane attractive potential. 
We expect $m, r_0$ to be approximately the string scale. Then, if $c \sim 10^{-2}$
the attractive force will be negligible. The well-known logarithmic attractive potential has been previously pointed out in the case of axion monodromy inflation in \cite{Conlon:2011qp}.

We will not aim for a greater degree of precision than the above arguments, and assume that there is a region in configuration space where the destabilization is small. From the perspective of the string landscape, this makes sense; by scanning over fluxes, one can explore all regions of configuation space, and the vacuum solutions which are in the regime of instability are discarded.

Branes wrapping the corresponding cycle of the axions can induce  
monodromies, which leads to a mechanism realizing theoretically self-consistent, large field, slow-roll inflation (see \cite{McAllister:2014mpa} and references within).
Let us consider the potential generated by axion monodromy, again focusing on RR axions descending from $C_2$.  For these axions the symmetry can be broken 
by considering a NS5-brane with two directions wrapping the two cycle $\Sigma$ associated with the axion. The resulting potential is given by the Born-Infeld action  \cite{McAllister:2008hb}
\be \label{potential1}
V(c_a)=\frac{\epsilon_{warp}}{(2\pi)^5 g^2_s \alpha^{\prime \; 2}} \sqrt{l^4 +(2 \pi g_s c_a)^2},
\ee
where $\epsilon_{warp}$ captures the possible effects of warping, $l\sqrt{\alpha^\prime}$ is the size of the two-cycle and we see for $c_a \gg l^2/(2 \pi g_s )$ the potential is linear in $c_a$ -- the shift
symmetry has been broken by the presence of the wrapped brane leading to a linear potential for the axion. 

In addition to the monodromy effect, D-brane and world sheet instantons 
can break the shift symmetry of the axion.
Such corrections should be generically expected and imply a contribution to the potential
\be \label{V_osc}
\Delta V(\chi_a) = \sum_i \Lambda_i^4 \cos(\chi_a/ f_a),
\ee
where we introduce the canonically normalized field $\chi_a = c_a f_a$ and we must sum over all such contributions that give a significant contribution to the potential.  
These contributions break the continuous shift symmetries to discrete ones with $\chi_a\rightarrow \chi_a +2\pi f_a$, and such terms can lead to additional important contributions 
to the potential and so must be checked against \eqref{potential1} and the slow-roll conditions.

We now turn to the question of $D3$ charge and tadpole cancellation in these models. The induced $D3$ charge on the $NS5$-brane is given by 
\be \label{inducedD3}
N_{D3, induced} \, = \, \frac{1}{(2\pi)^2 \alpha^{\prime}} \int_{\Sigma} C_2 \,\,.
\ee
This quantity turns out to be given by $N_{D3, induced} = \frac{\phi}{2\pi f}$, where $\phi$ denotes a generic field that may be the inflaton. Thus, $N_{D3, induced}$ can be quite large in the case when $\phi$ is the inflaton and $\frac{\phi}{f} \, \gg \, 1$. 

Suppressing the energy density of the wrapped brane to match observations will force us to place the branes in warped throats. Thus, there is a $D3$ charge contribution coming from the warping. We will denote the $D3$ charge of the throat by $N_{D3, throat}$. The total $D3$ charge of the system is then given by
\be
N_{D3, total} \, = \, N_{D3, induced} + N_{D3, throat} \,\,.
\ee
This will be cancelled by the orientifold action that wraps an anti-5-brane on the cycle $\Sigma^{\prime}$ in a throat with anti-$D3$ charge given by $-N_{D3, throat}$.

\section{A String Model of Gauge Field Production \label{stringy}}

In the last Section, we have gathered all the tools required for our construction. In this Section, we will give a string construction that realizes the model of gravity wave production discussed in Section \ref{section4}.

We will begin by sketching how such a setup is achieved within the string compactification and the model building constraints that result. Crucial to gauge field production (for the purposes of our model, and more generally for realistic reheating in these classes of models) will be the introduction of magnetized 5-branes in the $CY_3$ geometry. In the next subsection, we will discuss this topic. We will then describe the inflaton and spectator field dynamics in terms of UV data, and give the UV constraints that appear in our construction. 

Recall from Section \ref{section4} that
we are interested in an inflationary sector that successfully provides at least $60$ e-folds of inflation, coupled only gravitationally to spectator fields.
The action was 
\be \label{ole_action}
S= \int d^4x \sqrt{-g}~\left[ \frac{1}{2} m_p^2 R - \frac{1}{2} (\partial \vp)^2 - V(\vp)  + \mathcal{L}_{hidden}[\partial_{\mu}\chi,\chi ,F]\right],
\ee
\be \label{hidden_sector_L}
\mathcal{L}_{hidden}= -\frac{1}{2} (\partial \chi)^2 - U(\chi)-\fr{1}{4}F_{\mu\nu}F^{\mu\nu}-\fr{\chi}{4f}F_{\mu\nu}\tilde{F}^{\mu\nu},
\ee 
where $\varphi$ was the inflaton and $\chi$ and $F_{\mu \nu}$ were spectator fields.

Given both the axion decay constant \eqref{decay_constants} and potential energy \eqref{potential1} and ensuring that the oscillating contribution \eqref{V_osc} is subdominant, we can construct 
a slow-roll inflation model as done in \cite{McAllister:2008hb,Flauger:2009ab}. 

\begin{figure}[t!]
\begin{center}
\includegraphics[scale=0.4]{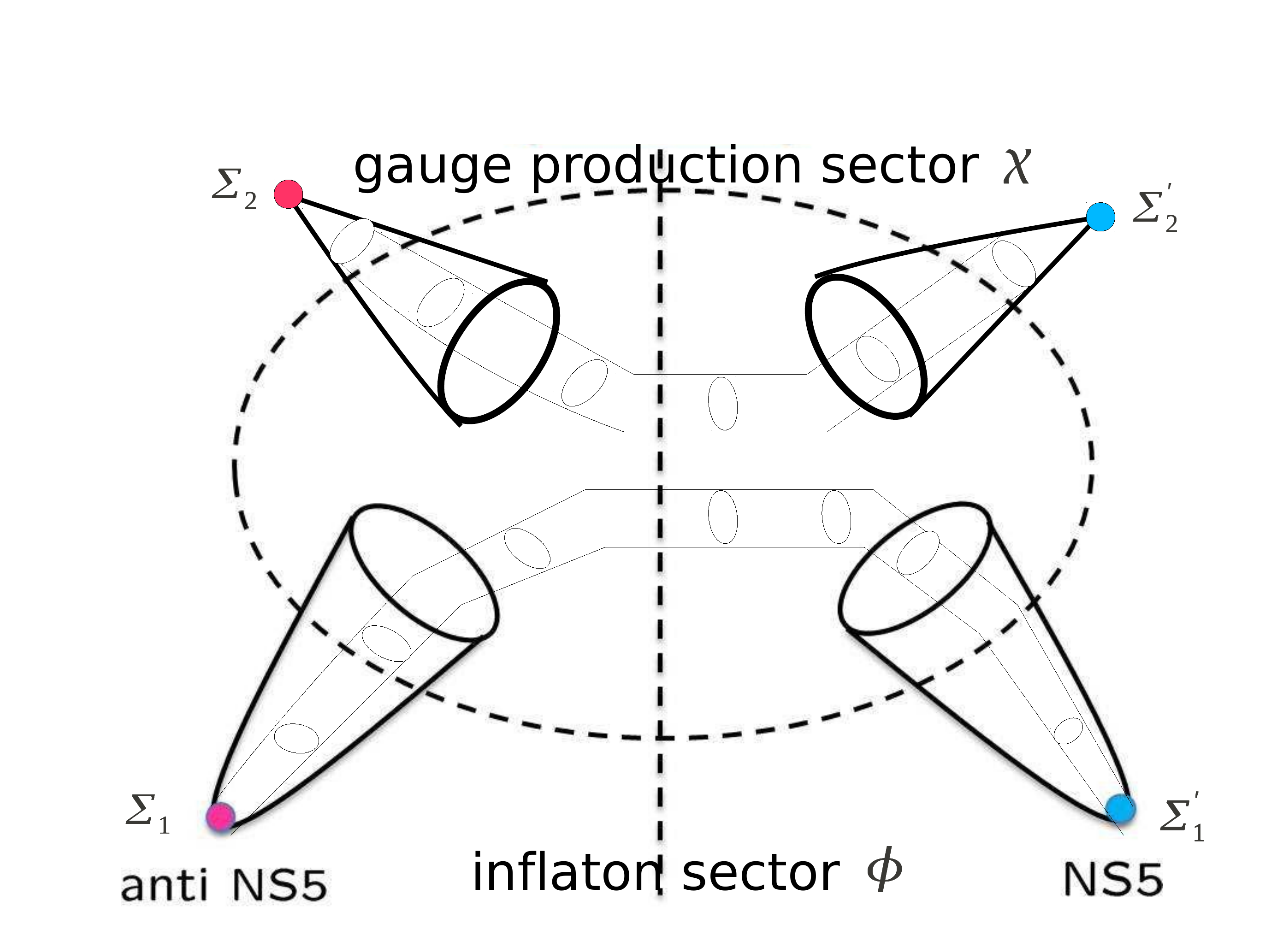}
\end{center}
\caption{A cartoon of the model for gauge field production from a sector $\chi$ that is gravitationally coupled to the inflaton $\phi$.} \label{cartoon}
\end{figure}

A sketch of our construction is provided in Figure \ref{cartoon}. We will be interested in a compactification with 
\be
h^{1,1}_- \,\, \geq \,\, 2 \,\,.
\ee
The inflaton sector is engineered by wrapping an $NS5$-brane on a cycle $\Sigma_1$ that supports the axion $\phi$. Under the orientifold action, an anti-$NS5$-brane wraps a cycle $\Sigma_1^{\prime}$ for tadpole cancellation. 

We note that placing the cycles $\Sigma_1$ and $\Sigma_1^{\prime}$ in warped throats is necessary to suppress the inflationary potential energy coming from the wrapped $NS5$-brane and match the COBE normalization. The curves $\Sigma_1$ and $\Sigma_1^{\prime}$ are thus members of a one parameter family of holomorphic curves $\mathcal{E}_1$ which extends down a warped throat (we denote this throat by T1). 

Similarly, the sector $\chi$ responsible for gauge field production is engineered on cycles $\Sigma_2$ and $\Sigma_2^{\prime}$ that are members of a one-parameter family of holomorphic curves $\mathcal{E}_2$. This family extends down a different warped throat T2, and we assume that   $\mathcal{E}_1$ and  $\mathcal{E}_2$ do not intersect. This enables our system to satisfy the requirement that the two sectors are only gravitationally coupled. One can ask how generic such a set-up is. Such multi-axion models that are coupled only gravitationally can be generically anticipated given the large number of two-cycles within a typical $CY_3$ geometry. Indeed, this fact helps motivate the notion that monodromy inflation could proceed via two or more axions -- an idea first pursued in \cite{Berg:2009tg}. 

We need to couple the $\chi$ field to gauge fields in our low energy Lagrangian. Given a gauge field $F$ on the brane, the desired operator descends from a Chern-Simon's term 
\be
\int C_2 \wedge F \wedge F
\ee
upon compactification to $4$D \cite{Svrcek:2006yi}.  In the context of axion monodromy model building such a term has already been considered for; gauge field production associated with the inflaton sector\footnote{In \cite{Barnaby:2011qe} the authors considered gauge field production in the directly-coupled-to-inflaton case, which as discussed in Section \ref{direct_coupling_section} is in tension with existing bounds on non-Gaussianity from Planck.} \cite{Barnaby:2011qe}, as a way to reheat at the end of inflation \cite{Blumenhagen:2014gta}, and as a possible constraint from CMB rotation when the axion corresponds to a quintessence field \cite{Panda:2010uq}.  

We note that phenomenological requirements such as a visible sector with chiral matter (and/or obtaining acceptable reheating into visible sector fields) will in general require us to turn on fluxes on the branes supporting the inflaton and the gravitationally coupled field $\chi$. To illustrate this, we consider an example where $\chi$ couples to an appropriately constructed visible sector, and turn on magnetic flux on the 5-branes wrapping $\Sigma_2$ and $\Sigma_2^{\prime}$. We will prefer to turn on an effective $D3$-brane charge $p$:
$$ 
{1\over 2\pi} \int_{\Sigma} F = p-1 \,\,.
$$ 
It is important to perform a number of checks when introducing brane fluxes into the system. We now turn to a discussion of the subtleties that arise.

\subsection{Magnetized 5-Branes in Orientifold Compactifications}

Magnetized branes in type IIB have a long history in the context of building semirealistic string vacua with chiral matter. These models are $T$-dual to models of intersecting $D6$-branes in type IIA. Chiral matter is vital in any phenomenological construction of low-energy physics, as well as any realistic reheating model based on axion monodromy inflation. We refer to \cite{Blumenhagen:2005mu} for a review of the $T$-dual intersecting brane models, and to \cite{Blumenhagen:2003vr} for the type IIB picture in toroidal models \footnote{We also refer to \cite{Cascales:2003zp}, \cite{Font:2004cx}, \cite{Cvetic:2005bn}, \cite{Marchesano:2004xz} for a sample of the literature on toroidal magnetized brane models.}.  

For our purposes, we will require magnetized 5-branes in $CY_3$, going beyond the simple toroidal picture. We will mainly follow \cite{Diaconescu:2006nk} in our treatment. Although we will be describing everything in the language of $D5$-branes, it can be adapted to the case of $NS$-branes as well.

Starting with the $N=2$ theory, it is clear that the system breaks tree level supersymmetry since the brane and anti-brane preserve different fractions of the bulk $N=2$. The $N=1$ supersymmetry preserved is determined by the central charge of a brane. For the brane and anti-brane, the central charges are respectively
\be
Z_+ = Z_{D5} + pZ_{D3}, \,\,\,\,\,\, Z_- = -Z_{D5} + pZ_{D3} \,\,\,.
\ee
The phases of $Z_+$ and $Z_-$ are not aligned for any deformation of the bulk Kahler structure away from the $Z_{D5} = 0$ locus, leading to breaking of brane world-volume supersymmetry. This breaking can be described by supergravity $D$-terms at weak string coupling and in a small neighborhood of the marginal stability locus $Z_{D5} = 0$ in the Kahler moduli space \footnote{For large deformations from the locus of marginal stability, string field theory computations are required to accurately calculate the brane dynamics.}. The Fayet-Iliopoulos couplings in the low energy gauge theory have been used to construct de-Sitter vacua in for example \cite{Burgess:2003ic}, where background fluxes and non-perturbative superpotential contributions fix moduli, followed by an uplift utilizing $D$-terms from $D7$-brane fluxes.
 
We note that the supersymmetry breaking is best parametrized by the parameter $\theta$ given by the relative phase between the charge $Z_+ = Z_{D5} + p Z_{D3}$ and $Z_{D3}$
\be
\theta \, = \, \frac{1}{\pi}(Im \ln{Z_+} - Im \ln{Z_{D3}}) \,\,.
\ee
The parameter $\theta$ has a minimum at the Landau-Ginzburg point in the non-geometric phase \cite{Diaconescu:2006nk}.

In our case, we will not be interested in studying supersymmetry breaking effects. We will remain strictly in the geometric phase for our model for the gauge field production mechanism we are interested in. The supersymmetry breaking induced by the inflaton sector will in any case be more dominant than $D$-term contributions from brane flux. However, in a detailed reheating model after the end of inflation based on this class of models, it would be necessary to take into account any supersymmetry breaking effects carefully. It may be useful then to stay near the large complex structure limit in the complex structure moduli space, and utilize mirror symmetry to situate the calculations in the supergravity limit of the mirror type IIA. We leave these considerations for future work. 

As noted before, there is a small subtlety in the brane-anti-brane stability issues we had studied earlier. The dynamics of the brane are encoded by fluctuations of the embedding map $i: \Sigma \hookrightarrow X$, which are described by sections $\zeta$ of the normal bundle $N_{\Sigma/X}$ of $\Sigma_+$. As described in \eqref{braneantibranepot}, a stable system can generally be obtained. However, according to the $\Pi$-stability analysis of \cite{Diaconescu:2006nk}, there is a tachyonic contribution to the mass of the lightest open string modes between the brane-antibrane pair that is proportional to $\theta$. Since the curves $\Sigma_2$ and $\Sigma_2^{\prime}$ are isolated, the positive contribution to the mass of the open strings should be typically much larger than the tachyonic contribution $\theta$. 

We now turn to the issue of moduli stabilization and superpotential terms in the presence of branes with fluxes. In the bulk $CY_3$, we will be utilizing a usual background RR flux compactification \cite{Giddings:2001yu}, \cite{Kachru:2003aw}. However, we now have to take into account the superpotential contribution from the magnetized brane. Brane-flux superpotentials have been studied in \cite{Lerche:2002yw}, \cite{Lerche:2002ck}, and in the context of $F$-theory in \cite{Grimm:2009ef}. For $D5$-branes, we refer to the detailed work of \cite{Grimm:2008dq} and references therein. Here, we summarize the essential points that are relevant for us. The main point is that at the level of the superpotential, one should use a combined brane-RR flux superpotential given by
\be \label{branefluxW}
W \, = \, \int_{X} F_3 \wedge \Omega + W_{D5} \,\,,
\ee
where $W_{D5}$ is in general a function of the deformation moduli $\zeta$, the complex structure moduli of the $CY_3$, and the brane flux $F$ \footnote{We note that the superpotential is not generally separable in the form given in \eqref{branefluxW}. This is possible in special cases, for example if $X$ contains a connected family of holomorphic curves interpolating between $\Sigma$ and $\Sigma^{\prime}$. For simplicity, we will assume that this is the case, and refer to \cite{Lerche:2002yw} for more details.}. For our purposes, we will assume that tuning the background RR-fluxes and brane flux appropriately in the combined superpotential $W$ will stabilize the complex structure moduli.

Before moving on, we mention several other caveats to this analysis. We have not explicitly discussed the case of turning on flux on the cycles $\Sigma_1$ supporting the inflaton, as would ostensibly be required for reheating. It would be interesting to compute the effect of brane flux on the slow-roll conditions. Moreover, our inflaton sector is present in a warped throat with background $D3$ charge. Generally, the magnetized $D5$ and background $D3$ should attract each other and the system should decay to a state where the $D3$ has been converted to magnetic flux on the $D5$. It would be interesting to compute the relevant stability conditions.

In the next subsection, we discuss the microscopic parameters in the inflaton and $\chi$ sector potentials. We then go on to a discussion of the consistency of the construction.

\subsection{Microscopic Parameters in the Inflaton and Hidden Sector}
In this subsection, we write down the potential for the inflaton and $\chi$ sectors using the microscopic data we have developed till now.

The action of the inflationary sector is
\be \label{inflate_act}
\int d^4x \sqrt{g} \left( \frac{1}{2} (\partial \varphi)^2 - \mu^3_\varphi \varphi \right) + \mbox{corrections},
\ee
where we have labeled the canonical inflaton arising from the RR axion as $\varphi = f_\varphi c$ with the corresponding decay constant following from \eqref{decay_constants}
\be
\left( \frac{f_\varphi}{m_p}\right)^2 =\frac{g_s}{8 \pi^2} \left( \frac{c_{\alpha \Sigma_1 \Sigma_1} v^\alpha }{{\cal V}_E}\right)
\ee
with the sum running over the remaining two-cycles of the compactification. 
Expressing \eqref{potential1} in terms of the canonical field $\varphi$ the parameter $\mu$ is then
\be \label{muparam}
\mu^3_\varphi = \frac{\epsilon_{(warp, \varphi)}}{g_s(2\pi \sqrt{\alpha^{\prime}})^4 f_{\varphi}},
\ee
As shown in \cite{McAllister:2008hb} adequate inflation and accounting for the COBE normalization requires $\mu \simeq 6.4 \times 10^{-4}$. Warping in the throat T1 can enable enough suppression of the inflationary energy to match observations.

Given the inflationary sector, we now turn to the spectator field ($\chi$) responsible for the gauge field production.  Recall from Section \ref{section4} that we require this field to be slow-rolling as well. In our construction, we achieve this by introducing an additional axion wrapped around a cycle $\Sigma_2$ belonging to the non-intersecting family $\mathcal{E}_2$. The monodromy term, in conjunction with an appropriate choice of the warping in throat T2, can then be used to ensure that this field is $(a)$ slow-rolling meeting the model building requirement that $\epsilon_\chi \ll \epsilon \ll 1$ and $(b)$ sub-dominant in energy density to the inflaton sector \footnote{Strong warping can lead to additional corrections to the monodromy potential \cite{Panda:2010uq}, as well as altering expressions like the Planck mass in \eqref{dimreducedplanck}.  Here we will be interested in {\em mild} warping and work in the approximation utilized in \cite{Kachru:2003sx}, where it was shown that in the region of interest warping effects can be safely ignored.}.

Denoting the warp factor of the spectator by $\epsilon_{(warp, \chi)}$, and keeping in mind that we require $\epsilon_{(warp, \chi)} \ll \epsilon_{(warp, \varphi)}$, the spectator sector is then specified by the decay constant
\be
\left( \frac{f_\chi}{m_p}\right)^2 =\frac{g_s}{8 \pi^2} \left( \frac{c_{\alpha \Sigma_2 \Sigma_2} v^\alpha }{{\cal V}_E}\right)
\ee
with the sum running over the remaining two-cycles of the compactification, 
and the potential from \eqref{potential1} is
\be \label{muparam_chi}
\mu_\chi^3 = \frac{\epsilon_{(warp, \chi)}}{g_s(2\pi \sqrt{\alpha^{\prime}})^4 f_\chi},
\ee

The low-energy description is completed once the brane flux $F$ is turned on. The Chern Simon's term connects $\chi$ and $F$, with an action given by
\be
S_{gauge}=\int d^4x \sqrt{-g} \left[ -\frac{1}{4} F^2 - \frac{1}{4} \alpha_{\mbox{\tiny brane}} \chi F_{\mu \nu} \tilde{F}^{\mu \nu}  \right]
\ee
The coupling $\alpha$ is given by
\bea
\alpha_{NS5}&=& \frac{C_0g_s^2}{(2\pi)^2 f_\chi}, \\
\alpha_{D5}&=&\frac{2 \pi g_s^{1/2}}{v_{\Sigma_2} f_\chi},
\eea
depending on whether we use $D5$ or $NS5$ branes. 

We note that the coupling matches the notation of \cite{Svrcek:2006yi} up to factors of $2 \pi$ coming from how the four dimensional gauge kinetic term is defined.

\subsection{Consistency Conditions}
In this subsection, we will take the most important conditions on the microscopic data required to build the specific model in Section  \ref{section4}. We must ensure that the low energy constraints outlined in Section \ref{section4} are satisfied, and also that the string construction is under control.

The microscopic data that determines the model by fixing the values of the quantities $(f_{\varphi}, f_{\chi}, \mu_{\varphi}, \mu_{\chi})$ in the low-energy Lagrangian is given by
\be
{\rm Microscopic} \,\,\,{\rm data:} \,\,\,\,\,\, (c_{\alpha \Sigma_1 \Sigma_1},\,\,\, c_{\alpha \Sigma_2 \Sigma_2}, \,\,\, \epsilon_{(warp, \varphi)}, \,\,\, \epsilon_{(warp, \chi)}, \,\,\,  v^\alpha, \,\,\, \mathcal{V}_E \,\,)
\ee
In the Appendix, we list the possible corrections to slow roll and back-reaction effects on moduli stabilization, and the methods employed in the literature to build consistent inflationary models in these scenarios. These conditions are not particularly specific to the model we are building; rather, they must generally hold in axion monodromy models in type IIB compactifications.
Before we proceed, lacking a full-fledged $CY_3$ construction, we will make the simplifying assumption that all intersection numbers satisfy
\be
c_{\alpha \Sigma_1 \Sigma_1}, \,\,\,\, c_{\alpha \Sigma_2 \Sigma_2} \,\,\, \sim \,\,\, \mathcal{O}(1) \,\,.
\ee
A statistical analysis following the work of Kreuzer-Skarke is also possible (we refer to \cite{He:2013epn} for an accessible recent review).

The first condition we require is that the $\chi$ sector energy density is sub-dominant to the inflaton sector. This can be satisfied by choosing
\be
\epsilon_{(warp, \chi)} \, \ll \, \epsilon_{(warp, \varphi)}
\ee
The second condition that we found in Section \ref{section4} is that the axion decay constant $f_\chi$ should lie in the range given by \eqref{fconstrain2}. As it turns out, the condition on the axion decay constant is intimately connected to the question of keeping $\alpha^{\prime}$  corrections under control, so that we remain in the supergravity regime as outlined in Section \ref{orintdataax}. This condition reduces to keeping the volume of the $CY_3$, and in particular, the volumes of the two-cycles, large enough to remain in the geometric regime.

The issue is not only one of computability in the geometric regime. In principle, one has to check whether $D5$ and $D3$ branes are stable BPS states in the non-geometric regime. The stability of BPS states in non-compact $CY_3$ has been studied in \cite{Douglas:2000qw}, \cite{bridge1}, \cite{bridge2}. The situation is less clear in compact $CY_3$ manifolds. We will avoid this problem by remaining close to the large radius limit. 
This condition can be placed in a number of ways. For example, following the classic work of \cite{Becker:2002nn}, the $\alpha^{\prime}$ corrections to the volume in the Einstein frame can be obtained in terms of the Hodge numbers of the $CY_3$, which will translate into a condition on $f_{\chi}$. We will find it more useful to use the approximation outlined in \cite{Flauger:2009ab}. Controlling worldsheet instanton corrections, the limit obtained is
\be
v^{\alpha} \, > \, \frac{1}{\pi \sqrt{g_s}} \,\,.
\ee
This will give a lower bound on $f_{\chi}$, thus giving us
\be \label{finalcondonf}
\frac{g^{1/4}_s}{(2\pi)^{3/2}\sqrt{\mathcal{V}_E}} \, < \, f_{\chi}/m_{pl} \, < \, 1 \,\,.
\ee

We note that there is a constraint on $f_{\chi}$ coming from requiring that the induced $D3$-brane charge $N_{D3, induced}$ (which depends on $f_{\chi}$ through \eqref{inducedD3}) be small enough to keep our model local and not distort the throat geometry. This is outlined in the Appendix. However, this condition is milder than \eqref{finalcondonf}.

To agree with the lower bound in \eqref{fconstrain2}, we require from \eqref{finalcondonf}
\be
\mathcal{V}_E \, > \, 10^6 \cdot \frac{g^{1/2}_s}{(2\pi)^3} \,\,.
\ee
Taking  taking $g_s \sim 0.1$, one obtains $\mathcal{V}_E \, \gsim \, 1000$.
Even without going into the details of moduli stabilization, it is clear that this is a sensible condition which should be easy to satisfy in a typical compactification. We thus reach the conclusion that there is no general obstacle to realizing the model in a string construction.

\section{Conclusions}
In this paper we have considered whether particle production and non-adiabaticity during inflation can lead to a competitive source of primordial gravity waves during inflation.  In all of the examples we considered, we found that even when these events lead to a detectable level of B-modes that the scale of inflation must be quite high. Stated another way, polarization observations would still be teaching us about the scale of inflation.  We identified the most promising case as models where the spectator fields are gravitationally coupled.  We then considered the UV completion of these models in the context of Type IIB flux compactifications with Axion Monodromy.   
The embedding served two purposes.  Firstly, although we saw that the range in which the axion decay constant leads to phenomenologically interesting results is quite narrow, there seems to be no obvious obstruction to realizing the setup in string theory.  Indeed, for weakly coupled constructions the primary model building requirements are mild warping and requiring the overall volume to be about a $1000$ times the string scale -- both conditions easily accommodated in typical compactifications.  Secondly, through the UV completion we have argued that it is possible to suppress dangerous sinusoidal terms that are known to spoil the gauge field production.   Although a more detailed investigation is needed, this provides further support for the gauge production models that have been considered in the literature. As we note in the text, these models could also provide an interesting approach to reheating at the end of inflation in models of Axion Monodromy.  

In summary, we find that although there can be competitive sources to the quasi-deSitter background for the origin of primordial B-modes, it seems challenging to vastly separate the scale of inflation from that implied by CMB polarization measurements.  In other words -- yes! -- we can really determine the scale of inflation.

\section*{Acknowledgements}
We would like to thank Joseph Conlon, Eugene Lim, Liam Mcallister, Eva Silverstein and Lorenzo Sorbo for useful discussions.  We would especially like to thank Sera Cremonini for initial collaboration and many useful discussions. This work was supported in part by NASA Astrophysics Theory Grant NNH12ZDA001N, and DOE grant DE-FG02-85ER40237. SW would also like to thank the DAMTP, Cambridge University and the Mitchell Institute for Fundamental Physics and Astronomy for hospitality.


\section*{Appendix A: Microscopic Conditions for Slow Roll and Back Reaction}

In this Section, we list the conditions for making sure that the slow roll potential is not ruined during inflation, or that moduli stabilization is not lost due to back-reaction effects. We also list the methods in which these issues are solved in the literature.

\bi
\item {\bf Possible Correction}: Destabilization of moduli during inflation. \\
 {\bf Resolution}:   For the RR axion the shift in the moduli potential during inflation was shown to be negligible as long as one requires $1 \gg v_+ \gg cg_s$, where
 $v_+$ is the two-cycle volume, $c$ is the axion and $g_s$ is the string coupling \cite{McAllister:2008hb}. \\
\item {\bf Possible Correction}: Backreaction of the NS5 brane on the Geometry and Renormalization of Planck Mass by new light species. \\
 {\bf Resolution}:   Wrapped NS5 on a two-cycle induces an effect D3-brane charge $N_{D3}$.  To avoid back-reaction we require $N_{D3} \ll R_\perp^4 / (4 \pi g_s \alpha^{\prime\,2})$ where $R_\perp$ is the smallest curvature radius transverse to the brane -- this is easily satisfied \cite{McAllister:2008hb}. \\
 \item {\bf Possible Correction}: Moduli Stabilizing Fluxes can generate potential for the axions. \\
 {\bf Resolution}: GKP \cite{Giddings:2001yu} orientifold stabilization in warped Type IIB use imaginary self-dual flux, which do not contribute to the axion potential. 
\item {\bf Possible Correction}: Inflation could destabilize moduli resulting in run-away to weak coupling or large-volume (Kallosh-Linde problem \cite{Kallosh:2004yh}). \\
 {\bf Resolution}: Focusing on the RR axion allows for $V_{inf} < V_{moduli}$ where $V_{moduli}$ sets the height of the barrier for escape.  Also, in \cite{McAllister:2008hb} it was demonstrated that shifts 
 in the moduli from inflation are also benign -- this is not the case for the NS axion $b$, which suffers an $\eta$ problem.  
\item {\bf Possible Correction}: Non-perturbative stabilization of Kahler moduli implies an $\eta$ problem for $b$-type axions since they mix with each other due to the appearance of $b$ in the 
Kahler potential. \\
 {\bf Resolution}: One can focus on $c$-type axions which do not mix with the volume, alternatively one could use perturbative methods to stabilized the volume \cite{McAllister:2008hb}.
\item  {\bf Possible Correction}: Moduli stabilization and Euclidean $D$ brane instanton corrections to the Kahler potential.  \\
 {\bf Resolution}: Exponentially suppressed by the size of the two-cycles $v_\alpha$ if taken larger than string scale.
\item  {\bf Possible Correction}: Moduli stabilization and Euclidean $D$ brane instanton corrections to the Super potential.  \\
 {\bf Resolution}: One can focus on stabilization of the volume using gaugino condensation on D7 branes.  The combined holomorphy of the gauge coupling and the super potential imply 
 that the instanton corrections are exponentially suppressed by the four-cycle volume. 
\item  {\bf Possible Correction}: NS$5$ brane wrapped on a two-cycle induces a tadpole through an effective $D3$ brane charge.  \\ 
{\bf Resolution}: Introduce $\overline{D3}$ on a nearby two-cycle to cancel (where `nearby' means at a distance small compared to the $D7$ used to stabilize the Volume via KKLT.
\ei

\bibliographystyle{utphys}
\providecommand{\href}[2]{#2}\begingroup\raggedright
\endgroup
\end{document}